\documentstyle[12pt,epsfig,here,rotate,osamat]{retema}
\renewcommand{\baselinestretch}{1.5}
\renewcommand{\arraystretch}{1.5}
\begin{document}
\newcommand{\beq}[1]{\begin{equation}\label{#1}}
\newcommand{\eeq}{\end{equation}}
\newcommand{\bsq}[1]{\begin{subequations}\label{#1}}
\newcommand{\esq}{\end{subequations}}
\newcommand{\bfn}[1]{{\bf#1}}
\newcommand{\gl}[1]{eqn. (\ref{#1})}
\newcommand{\gls}[2]{eqns. (\ref{#1}) and (\ref{#2})}
\newcommand{\fur}{\qquad\mbox{for }\,}
\newcommand{\taud}{$\tau_{\scriptscriptstyle D}$\ }
\newcommand{\mtaud}{\tau_{\scriptscriptstyle D} }
\newcommand{\taur}{$\tau^{\scriptscriptstyle R}$\ }
\newcommand{\mtaur}{\tau^{\scriptscriptstyle R} }
\newcommand{\taue}{$\tau^\varepsilon$\ }
\newcommand{\mtaue}{\tau^\varepsilon }
\newcommand{\taurr}{$\tau^{\scriptscriptstyle RR}$\ }
\newcommand{\mtaurr}{\tau^{\scriptscriptstyle RR} }
\newcommand{\fraq}[2]{{\displaystyle \frac{#1}{#2}}}
\title{Polymer--Mode--Coupling Theory of Finite--Size--Fluctuation Effects in
Entangled Solutions, Melts and Gels. I. General Formulation and Predictions}
\author{Matthias Fuchs$^+$
and Kenneth S. Schweizer\\
Departments of Materials Science and Engineering,\\ Chemistry,
and Materials Research Laboratory,\\
University of Illinois, 
1304 West Green Street, Urbana, Illinois 61801}

\date{May 1997}
\maketitle
\begin{abstract}
The transport coefficients of dense polymeric fluids are approximately
calculated from the microscopic intermolecular forces.
The following finite molecular weight effects are discussed within the 
Polymer--Mode--Coupling theory (PMC)
and compared to  the corresponding reptation/ tube ideas:
constraint release mechanism, spatial inhomogeneity of the entanglement
constraints, and tracer polymer shape fluctuations.
The entanglement corrections to the single polymer Rouse dynamics are shown to
depend on molecular weight via the ratio $N/N_e$, where the entanglement degree
of polymerization, $N_e$, can be measured from the plateau shear modulus.
Two microscopically defined  non--universal
parameters, an entanglement strength $1/\alpha$ and a length scale ratio,
$\delta= \xi_\rho/b$, where $\xi_\rho$ 
and $b$ are  the density screening and  entanglement length respectively,
  are shown to determine the reduction of the entanglement
effects relative to the reptation--like asymptotes of PMC theory.
 Large finite size effects 
are predicted for reduced degrees of polymerization
up to $N/N_e\le10^3$. Effective
power law variations for intermediate $N/N_e$ 
of the viscosity, $\eta\sim N^x$, and the diffusion constant, $D\sim N^{-y}$,
can be explained  with exponents significantly 
exceeding the asymptotic, reptation--like
values, $x\ge 3$ and $y\ge2$, respectively.
Extensions of the theory to treat tracer dielectric relaxation, and 
 polymer transport in gels and other amorphous
systems, are also presented.
\end{abstract}

\rightline{Macromolecules, accepted (May 1997).}

\noindent$^+$ Current address: Physik-Department, Technische Universit\"at
M\"unchen, D-85747 Garching, Germany

\newpage

\section{Introduction}

The transport properties of long chain polymers show characteristic phenomena 
attributed to chain ``entanglements'' \cite{ferry,degennes,de}. As these
effects are specific to macromolecules they appear connected to the internal
degrees of freedom of polymers.
 The reptation/ tube theory has developed an
approach to the problem of entanglements of flexible chain polymers, which has
proved very versatile, and has found wide spread use
\cite{degennes,de,degrep}. Essential for this approach is the postulate of a
phenomenological concept, the confining ``tube'' and anisotropic
reptative motion of a polymer in the tube. 

Recently, a microscopic theoretical approach has been formulated by one of us
\cite{kss1,kss2}. It starts from approximations to the exact expressions for
the microscopic forces and attempts to derive the dynamics of entangled
polymers from the underlying equilibrium structure of the polymeric liquid.
The connection of both theories has not been worked out comprehensively, as the
theoretical descriptions and some of the involved approximations strongly differ.
However, for linear chains in 3--dimensions
both theoretical approaches arrive at identical predictions for the
exponents characterizing the asymptotic scaling of the transport properties
with molecular weight. 

It is well documented that experimental studies of
entangled polymeric melts and solutions find only partial agreement of
predicted and measured exponents \cite{ferry,rotprag}. Especially, the 
long standing issue of the dependence of the shear viscosity on molecular
weight should be mentioned; experimentally $\eta\sim M^{3.4\pm0.2}$ is observed
\cite{ferry}, whereas both theories predict $\eta\sim M^3$ asymptotically.
Moreover, recent experiments on entangled polymer solutions
\cite{nemoto1,nemoto2,nemoto3}, and of the diffusion of polymer  tracer chains
in crosslinked gels \cite{lodgegel,hoagland}, find strong deviations from 
the theoretical  predictions for the diffusion coefficients,
 $D \sim M^{-2}$. Larger exponents, e.g. $D\sim
M^{-2.8}$, are reported in gels \cite{lodgegel} and solutions, whereas in
polymer melts the predicted exponents apparently describe the limiting behavior
to within experimental error 
\cite{greenkramerlet}.  At present the findings in gels and solutions
cannot be rationalized within the reptation theory \cite{nemoto} even including
non--asymptotic   corrections  resulting from the finite size of the tracer or
matrix polymers \cite{grassley,klein1,klein2,doicont1,doicont2}.
Another set of recent intriguing experiments are the measurements of the
dielectric relaxation times of tracer polymers in highly
entangled polymeric melts
\cite{adachi,adachi96}. Depending on the matrix molecular weight, Rouse,
$\tau\sim M^2$, reptation--like scaling, $\tau\sim M^3$, or power law
scaling with exponents in between 2 and 3 is observed. In all cases, except for
the Rouse limit, however, the distribution of relaxation rates characterizing
the dielectric disentanglement process is much broader than expected from the
reptation results for the tube survival function \cite{de}.

In this paper
we study whether the recent polymer mode coupling (PMC) theory
\cite{kss1,kss2},  which recovers
the reptation like scalings,  $\eta \sim N^3$ and $D\sim N^{-2}$, in the
asymptotic regime, can account for the observed different  
molecular weight dependences for 
finite $N$.  In particular the PMC description of the
constraint release mechanism   and the constraint porosity corrections are
analyzed within a  simplified model for polymer liquids.
Our goal is a unified understanding of {\it all} the puzzling non--asymptotic
behaviors, for tracer and self diffusion and chain relaxation in solutions,
melts and gels, within a single theoretical framework formulated at the
molecular level.  Throughout
this paper extensive comparison with the reptation/ tube approach is drawn in
order to explain the physical content of PMC theory. Tracer dynamical
shape fluctuations are also
included in our study since  they are the origin of power law frequency
behavior in the short time asymptote of the disentanglement process in 
the shear modulus or the end--to--end vector correlation function.  

In the accompanying paper \cite{pap2},  the theoretical
description will be tested in comparisons with various experimental data sets. 
A major  virtue of the PMC approach
which starts from the microscopic force balance
equations is to provide connections between 
the dynamics and the underlying equilibrium liquid and macromolecule 
structure. This
allows  independent theoretical or experimental information to be used
in order to
predict the magnitudes and trends of the finite size corrections
and of the asymptotic prefactors. These estimates,
 and discussions of both extensions 
of the pure reptation/ tube picture to include finite size corrections and 
 alternative non--reptation
 approaches to the dynamics of entangled polymers, will be
included in the second paper. 

The contents of this  paper is as follows.
Section 2 presents those aspects of the theory which are required for the
present 
discussion, and a careful enumeration of the  necessary approximations.
A new derivation of PMC theory based on the collective shear stress field as
a primary slow variable is also presented.
In section 3, the origins of finite size effects within the PMC approach
are discussed, 
and simple models for their description are introduced. The previously obtained
asymptotic predictions of PMC theory are  summarized in section
4.A.  Section 4.B develops
the  theoretical formulae including finite $N$ corrections
for the diffusion constants, viscosities, and the
 internal
and end--to--end--vector relaxation times.
In section 5, model calculations  for the transport coefficients,
tracer and self diffusion constants, viscosities and dielectric relaxation
times,  of polymeric melts and solutions show the importance of finite size
effects, and a discussion of their physical
origins within the PMC description is presented.
 Extensions of the theory to
polymer tracer diffusion through gels are presented and analyzed in section
6. The discussion in section 7
summarizes our findings and their connection to the reptation/ tube
approach. Finally, we note that the present and following papers are rather
long. We believe this is inevitable given our goals: ($i$) development of a
very  general theory,
 ($ii$) clear explanation of the physical and mathematical
content of PMC theory and its similarities and differences with respect to 
reptation/ tube and other phenomenological approaches, and ($iii$) unified,
comprehensive description of all non--asymptotic corrections to the transport
properties of melts, solutions, and gels.
\section{Theory}

\subsection{Generalized Langevin Equations}

Use of the exact Mori--Zwanzig formalism allows one to derive a generalized Langevin equation for
the dynamics of the segments of a tracer polymer chain in a polymeric liquid
\cite{kss1}.
In order to proceed, knowledge of the microscopic forces acting on the tracer
segments is required. The ideality
concept of Flory \cite{flory}, which states that chain macromolecules in a
dense, polymeric melt exhibit ideal, random--walk intramolecular correlations
simplifies the intramolecular forces.  
The limit of large degree of polymerization, $N$, of the tracer, 
is mathematically attractive since the
discrete nature of the polymer segments can be neglected \cite{de,rouse}.
Then the continuous Gaussian space curve 
description is generally assumed to be valid leading to ${\bf F}^{\rm intra}_\alpha(t) =
K_S \partial^2_\alpha {\bf R}_\alpha(t)$,
where $K_S$ is the entropic spring constant and ${\bf R}_\alpha(t)$ 
the position
vector of segment $\alpha$ on the tracer polymer. Time and segment length
derivatives are
abbreviated by $\partial_t = \frac{\partial}{\partial t}$ and
$\partial^2_\alpha=(\frac{\partial}{\partial_\alpha})^2$ respectively.
Extensions to incorporate the discrete structure of the chains or non--Gaussian
intramolecular correlations, like chain stiffness, have been studied
\cite{bz3,freed}, but are neglected in the present work. Our approach,
therefore,
is limited to dynamical processes acting on length scales large compared to
segmental sizes. Formulating this restriction as an inequality,
$b \gg \sigma$ is required, where $b$ denotes the entanglement length (to be
discussed below) and $\sigma$ is an effective, Gaussian--segment size
determined by the persistence length of the semiflexible polymer
\cite{flory}. 

 Far less is known about the time and space correlations of the
intermolecular forces exerted by
the surrounding polymeric matrix on the tracer. 
The rapid collective local
dynamical variables, including when appropriate the
solvent,  give
rise to an instantaneous, uniform friction described by the monomeric friction coefficient,
$\zeta_0$.
Yet unspecified, slow collective degrees of freedom lead to  a non--trivial
memory function $\Gamma_{\alpha\beta}(t)$, the autocorrelation 
function of the slow fluctuating intermolecular forces \cite{kss1} 
${\bf F}^Q_\alpha(t)$:
\beq{1}
\zeta_0 \partial_t\; {\bf R}_\alpha(t) - K_S
\partial_\alpha^2\; {\bf R}_\alpha(t)
+ 
\int_0^N d\beta\; \int_0^t dt'\; \Gamma_{\alpha\beta}(t-t')
\partial_{t'} {\bf R}_\beta(t') = {\bf F}^Q_\alpha(t)\; .
\eeq
\beq{2}
 K_S = \fraq{3k_BT}{\sigma^2} 
\eeq
\beq{3} 
\Gamma_{\alpha\beta}(t) = 
\fraq{1}{3k_BT} \langle {\bf F}^Q_\alpha(t) \cdot
{\bf F}^Q_\beta(0) \rangle \;, \qquad\mbox{where }\;  
\langle {\bf F}^Q_\alpha(t) \cdot {\bf R}_\beta(0) \rangle = 0
\eeq
In \gls{1}{3} the possible spatial anisotropy of the intermolecular forces on
the segment size length scale is neglected, and a spatially isotropic
 motion is assumed  \cite{kss1,kawasaki}. Whether such an isotropic
description is literally valid, or represents a kind of effective medium
description of cage--averaged dynamics, remains unresolved.
The topology of linear chains determines the boundary condition of vanishing distortion at the free 
chain ends \cite{rouse,de}:
\beq{4}
\partial_\alpha {\bf R}_\alpha(t) = 0 \fur \alpha = 0 , N\; .
\eeq
The initial values or equilibrium correlations again follow from the 
ideal Gaussian  intramolecular structure \cite{de,flory}:
\beq{5}
\langle ( {\bf R}_\alpha(0) - {\bf R}_\beta(0) )^2 \rangle = \sigma^2 \; |\alpha-\beta | \; ,
\eeq
Special effects of semiflexibility are neglected here and \gl{5}
defines the effective  segment size $\sigma$.

These equations have been solved and their physical predictions have been
discussed by Schweizer and 
coworkers within the PMC theory based on various approximations
\cite{kss1,kss2,kss3,kss4,ks1,ks2,ks3,ks4,fractal}.
In the present paper we will improve
upon one of the  approximations, namely the frozen matrix assumption \cite{kss1}, in order to
study the consequences on the probe dynamics of taking into account more realistically the
dynamics of the surrounding medium. We consider the tracer to be embedded in a 
liquid of $\nu$, chemically identical chain polymers of degree of
polymerization $P$. The polymer 
segment density, $\varrho_m$, therefore is $\varrho_m=P\nu/V$. 

The tracer diffusion coefficient is defined in the usual manner
 from the motion of the center--of--mass,
${\bf R}^{\rm CM}(t) = N^{-1} \int_0^N d\alpha {\bf R}_\alpha(t)$, in the
hydrodynamic, or long time, limit \cite{de,hansen}:
\beq{6}
\langle ( {\bf R}^{\rm CM}(t) - {\bf R}^{\rm CM}(0) )^2 \rangle \to 6 D t \fur t \to \infty
\eeq
From the previous analysis of the PMC equations it is known that they describe a dynamical process
which has been
favorably  compared to the entanglement dynamics as measured by diffusion
experiments, rheology, pulsed field gradient NMR,
and dielectric spectroscopy \cite{ks1,ks2,ks4,fractal,nmr}.
The theory predicts the existence of a strongly $N$ dependent terminal or
disentanglement time, \taud, which in the present context can be defined by
the final relaxation step in the internal, conformational dynamics  of the
tracer polymer \cite{kss2}: 
\beq{7}
\langle {\bf F}^{\rm intra}_\alpha(t) \cdot {\bf F}^{\rm intra}_\beta(0)
\rangle \to 
\langle {\bf F}^{\rm intra}_\alpha(0)
 \cdot {\bf F}^{\rm intra}_\beta(0) \rangle \; e^{-t/2\mtaud}\; \fur N \to
 \infty\, . 
\eeq
Note that \gl{7} holds for long times only (Markov regime). It
 is simplified in so far as the cut--off of the PMC effects
below a certain entanglement length scale, $b$, is not denoted explicitly,
and also small corrections to the global modes are neglected
\cite{kss2,ks2,fractal}.  
The theory also predicts the existence of an entanglement plateau,
$G_N = \varrho_m k_BT / N_e$,  in the shear modulus on intermediate times
\cite{kss2}. (The  theoretical predictions for $N_e$ are quoted below.)
This result depends on the generally accepted, but rigorously
unverified, assumption
that the measured collective 
shear stress is dominated by the incoherently added, single chain
contributions 
\cite{de,zawada}.
From this amplitude of the entanglement process in the shear stress, $G_N$, and
the final relaxation time \taud of \gl{7},
 follows the shear viscosity \cite{kss2,ks2}:
\beq{8}
\eta =  G_{\rm N} \mtaud + \eta^{\rm R}  = \fraq{\varrho_m k_BT}{N_e} \mtaud + \eta^{\rm R}
\eeq
In \gl{8}, a simple additivity assumption was used in order to describe the
crossover from the 
low molecular weight, Rouse \cite{de,rouse} result, $\eta^R=\varrho_m\zeta_0\sigma^2N/36$,
 to the high  molecular weight, asymptotic
PMC  result, $\eta^{\rm PMC}=G_N\mtaud$. This assumption is expected to be
reliable at 
rather high molecular weights and compares well with numerical solutions to the
full PMC equations including crossover effects \cite{ks2}. 

The calculation of the tracer diffusion coefficient, $D$, and the internal final relaxation time, 
\taud, can proceed from the exact starting equations (\ref{1}) to
(\ref{5}) only via approximations. 
Calculating the diffusion coefficient is somewhat
 complicated by the coupling of translational and rotational 
motions as captured in \gl{1} in general. (See below for a further discussion of this point.)
Following previous analysis we will neglect this coupling which arises from the end monomers of
the tracer polymer, i.e. the boundary condition, \gl{4}, appropriate for linear
chains. Naturally, it is  also absent for cyclic polymers. We
argue that the diffusion over distances large compared to the size of the
polymer, $\Delta r \gg R_g=\sigma\sqrt{N/6}$, 
is determined by the sum of all intermolecular forces acting uniformly on the center--of--mass of the
 probe \cite{fractal}.
Then the diffusion coefficient is determined from the memory function of the
sum of all intermolecular  forces.
Reassuringly, the result for $D$ we will get is a clear
extension of known expressions for the diffusion coefficients of atomic or
colloidal systems \cite{hansen,hessklein}.

The dynamics of the uniform friction exerted on the center--of--mass is 
unknown and can only be found with approximations. In the approach put forward
in refs. 5 and 6
 the uncontrolled but often surprisingly successful
mode--coupling approximation 
\cite{kss1,hansen,kawa,fixman,boonyip,gs,yip}
is used. It requires a physically motivated choice of the relevant, slow modes.
We assume that the dynamics of the center--of--mass friction is
dominated by
the collective density fluctuations of the matrix, $c_k(t)$,
\beq{9}
c_{\bf k}(t) = \sum_{m=1}^\nu \int_0^P d\gamma \; e^{i{\bf kR}^m_\gamma(t)}\; ,
\eeq
and the collective  monomer density of the tracer
\beq{10}
\varrho_{\bf k}(t) =  \int_0^N d\alpha \; e^{i{\bf kR}_\alpha(t)}\; .
\eeq
In \gl{9} the time dependent
 vector ${\bf R}^m_\gamma(t)$ is the position vector of the
segment $\gamma$ on the matrix polymer $m$.
The total intermolecular fluctuating forces exerted on the tracer
center--of--mass  are therefore described by a
memory function $\Sigma(t)$ of the form \cite{kss2,fractal}:
\beq{11}
\Sigma(t) = \fraq{1}{3\zeta_0k_BTN} \int_0^N d\alpha d\beta \langle {\bf F}^Q_\alpha(t) \cdot
{\bf F}_\beta(0) \rangle 
\approx \fraq{Vk_BT}{3\zeta_0N} \int \fraq{d^3k}{(2\pi)^3} 
|{\bf k} V_k^{\rm eff}|^2  \langle c_{-\bf k}^Q(t) \varrho_{\bf k}^Q(t) 
c_{\bf k}^Q \varrho_{-\bf k}^Q \rangle\; ,
\eeq
where $V$ is the volume, and
the approximation results from assuming that the four point correlation
function of the  
fluctuating forces is dominated by its overlap with the correlator of the pair variables
$\varrho_{\bf -k}(t) c_{\bf k}(t)$. The label 
$Q$ indicates that the 
projected dynamics 
controls the time evolution of  the fluctuating forces. The projector $Q$
formally achieves 
that there is no further, linear coupling of the  monomer coordinates to the fluctuating forces besides
the one described explicitly  \cite{hansen,boonyip}  in \gl{1}.
The vertex, i.e. the normalized 
overlap of the forces with the slow pair variables, can be calculated
\cite{kss1}  (when neglecting three point correlations, i.e. $Q=1$):
\beq{12}
{\bf k} V^{\rm eff}_k 
= \fraq{1}{NP\nu k_BT} \int_0^N d\alpha d\beta \fraq{\langle c_{-{\bf k}} \varrho^\beta_{\bf k}
{\bf F}^Q_\alpha \rangle}{\omega_k S_k} 
= \fraq{-i{\bf  k} \varrho_m h_k}{ P\nu \omega_k S_k}\fur Q=1\; .
\eeq
It is the normalized equilibrium correlation of a collective  tracer variable,
$\int_0^N d\alpha {\bf F}^Q_\alpha$, with the product of a collective matrix
variable, $c_k$, and another collective  tracer 
variable, $\int_0^N d\alpha \varrho^\alpha_{\bf k}$. Therefore it is natural that the
total intermolecular 
correlation function, $h_k$, arises. This aspect
also holds for atomic and colloidal
systems \cite{hessklein,gs}. Whereas the collective 
intramolecular structure factor,
$\omega_k = N^{-1} \langle \varrho_{-{\bf k}}(0) \varrho_{\bf k}(0) \rangle$,
describes a single polymer, the total intermolecular correlation function,
$h(r)=h_{\alpha\beta}(r)= g_{\alpha\beta}(r)-1$, is an inter--chain
site--site radial
 distribution function correlating segments on different macromolecules. 
 A peculiarity of
polymeric tracers is the long--ranged spatial dependence of $h(r)$ due to the
well  known correlation hole  \cite{degennes} 
effect, $h(r) \sim \frac{-1}{r} e^{-r\sqrt
2/R_g}$ for $r\gg \sigma$. The existence of this 
long--ranged correlation can be argued on general grounds as 
a universal consequence of chain connectivity and interchain excluded volume
forces. 
 This correlation extends to the tracer size, $R_g$. Note that if the liquid is
 taken as a random continuum, $g(r)=1$, then $h_k\to 0$ and all entanglement
 effects vanish.
In the polymer reference  interaction site model
(PRISM) of Curro and Schweizer the correlation hole is rigorously
recovered, but also the local 
structure is described realistically \cite{kcur1,kcur2,kcur3}.
 This is necessary in the present context
as the short--ranged excluded volume interactions
dominate the polymer liquid structure and also determine the magnitude of
the entanglement friction. 
PRISM therefore has been used to calculate the equilibrium
structural correlation functions entering into the vertices
\cite{kss1,kss2,ks1,ks2,ks3,ks4,fractal}. PRISM  theory
connects $h_k$ to the intra and collective structure factors, $\omega_k$ and
$S_k$ respectively, where $S_k=\omega_k + \varrho_m h_k$,  and 
to an effective short ranged pseudopotential, $c_k \approx -V_k/k_BT$, the 
site--site direct correlation function.
\beq{13}
h_k = c_k \omega_k S_k
\eeq

Although the vertex is now  known,  the dynamics of the  projected pair 
variables, $\varrho^Q_{\bf -k}(t) c^Q_{\bf k}(t)$ is not.
The approach put forward by Schweizer \cite{kss1,kss2}
uses the mode factorization approximation to break the average on the right
hand side of \gl{11},
and interprets
the projected single chain dynamics of the tracer as the 
one arising from a perturbative  short time theory, the renormalized Rouse (RR)
model (see section 2.B below) and the projected collective matrix dynamics 
as the full entangled dynamics.
\beq{12.5}  \langle c_{-\bf k}^Q(t) \varrho_{\bf k}^Q(t) 
c_{\bf k}^Q \varrho_{-\bf k}^Q \rangle \approx P \nu N \omega^{\rm RR}_k(t)
S_k(t)\; .
\eeq

\begin{figure}[H]
\centerline{\rotate[r]{\epsfysize=18.cm 
\epsffile{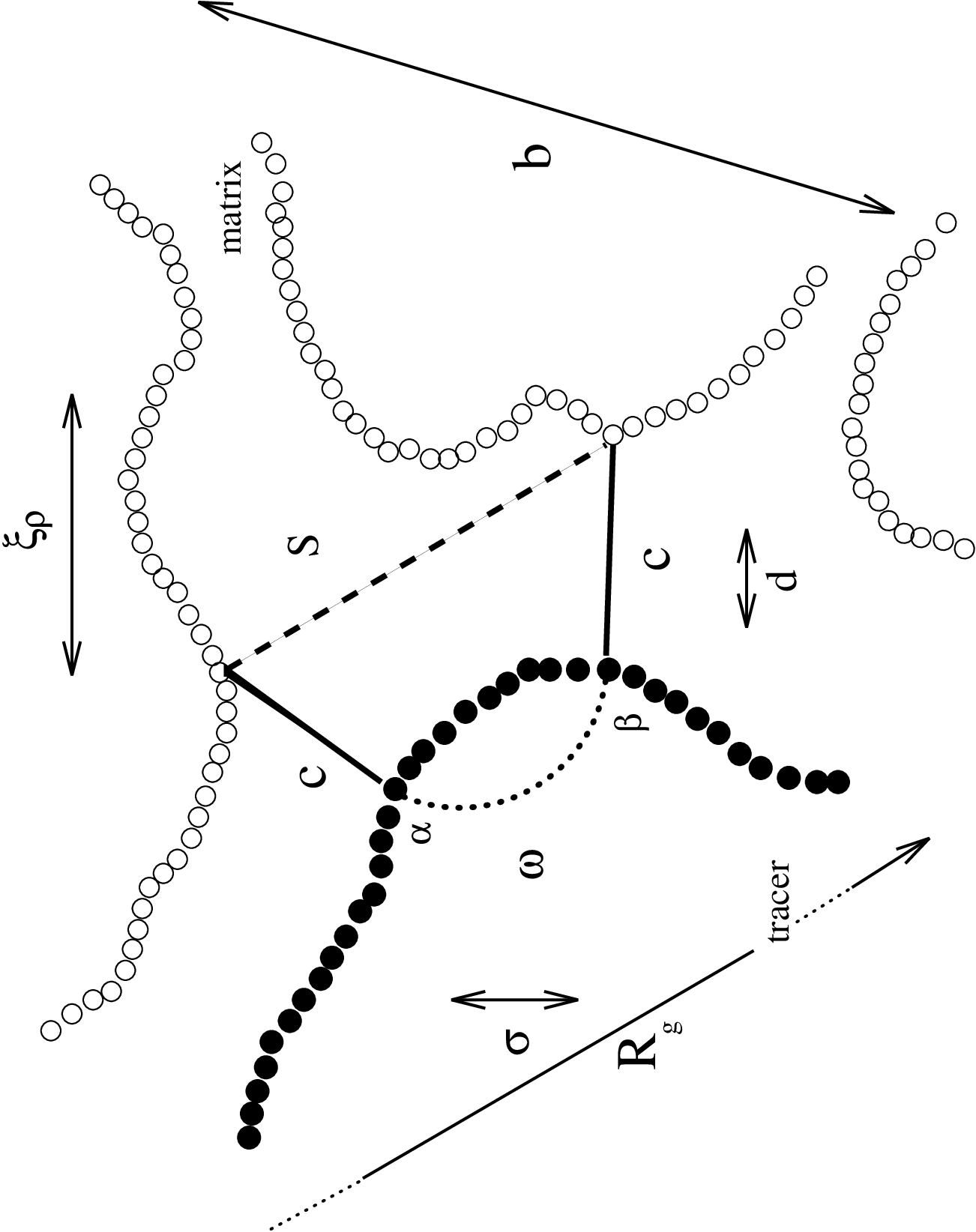}}}
\caption{ }\label{fsigm}
\end{figure}

Even though the intermolecular site--site correlation
 function, $h_k$, is long ranged, in the final result for
$\Sigma(t)$ the vertex of \gl{12} 
is relatively short ranged because of the anti-correlated normalization of
the projector.
\beq{14}
\Sigma(t) \approx \fraq{k_BT\varrho_m}{3\zeta_0} \int \fraq{d^3k}{(2\pi)^3} k^2 c_k^2
\omega^{\rm RR}_k(t) S_k(t)
\eeq
Eqn. (\ref{14}) expresses  that the major contributions to
the $t=0$ amplitudes of the
friction (fluctuating forces)
exerted on the tagged polymer
center--of--mass arise from short--ranged, local
intermolecular correlations. This can be seen from the wavevector dependence of
the friction contributions in \gl{14}.
Wavevectors large compared to $1/R_g$, corresponding to local distances,
are weighted most heavily as $k^4 \omega_k$ increases monotonically with $k$.
However, $kR_g\le 1$ long wavelength contributions still play a crucial role in
establishing the total Markovian friction due to their long relaxation times.
Figure \ref{fsigm} shows a schematic representation of the collective and
intramolecular correlations and length scales which determine 
$\Sigma(t)$. 
The tracer interacts with the matrix polymers via the short ranged, effective
potential, $c_k$, the direct correlation function. The spatial correlations of
the fluctuating, intermolecular forces further propagate along the tracer and
through the matrix medium. Intramolecular correlations along the tracer polymer
are described by $\omega_k$ and are characterized by two 
length scales, $\sigma$ (local) and $R_g$ (global). The density screening
length, $\xi_\rho$, and the entanglement length, $b$, characterize  the
constraining equilibrium and dynamic structure of the surrounding matrix.

The diffusion coefficient, $D$, can now be found in the Markovian, long time
limit of \gl{6},
\beq{15}
D^{-1} = \frac{\zeta_0 N}{k_BT} \, [ 1 + \hat \Sigma_0 ]
= \frac{\zeta_0 N}{k_BT}\, [ 1 + \int_0^\infty dt  \Sigma(t) ] \; .
\eeq
Extending previous work based on an effectively static, homogeneous matrix
($S_k(t)\approx S_0$), 
the present study focuses on the effects of the spatial
and time dependence of  the matrix fluctuations, $S_k(t)$ in \gl{14}, 
 on the magnitudes  and the molecular weight dependences
of the diffusion constant and other transport properties.

The calculations leading to the expressions for the internal relaxation time
\taud have been discussed 
in refs. 6, 31, and 34.
In order to show the robustness of these results, to provide further physical
interpretation, and as a basis for incorporation of all finite size
corrections, 
we re--derive them arguing differently. The original aim of Kawasaki when
using the mode--coupling 
approximation was to identify slowly varying, nonlinear contributions to the
memory functions and equations of motion
\cite{kawa,fixman,yip,hohenhalp}. 
One experimentally observed slow variable in polymeric liquids is the shear
stress. Its time scale, 
i.e. the shear disentanglement time, $\mtaud^s$, 
grows strongly with molecular weight. It is the longest time scale, at least
according to experimental observations, in entangled  polymeric liquids not
close to a demixing or order--disorder phase
transition \cite{ferry,degennes,de}. 
For example,  dielectric measurements
show that the internal, conformational relaxation time, $\mtaud^\varepsilon$,
very nearly equals $\mtaud^s$ in melts \cite{adachi}.
We  therefore assume the existence of one disentanglement time, \taud,
appearing in different collective and single chain properties
 which  originates in the single chain, conformational dynamics. Thus
\taud can  be calculated from the disentanglement
process of a (tracer) macromolecule for $N=P$ 
and is itself affected by the matrix
disentanglement process \cite{kss2} . 

As \taud is the longest relaxation time,
 any variable participating in the  disentanglement step can
be considered a slow mode.
Therefore, before we summarize the theoretical or
experimental evidence about the behavior of the matrix structure factor, $S_k(t)$, on this time scale,
let us calculate the fluctuating force memory function,
$\Gamma_{\alpha\beta}(t)$ of \gl{3},  assuming that it is
dominated by the collective stress tensor
which is known to be slow \cite{ferry}.
To that end, the steps in eqns. (\ref{11}) to (\ref{14}) will be repeated for
those intermolecular friction forces which affect the internal conformational
dynamics, replacing the collective matrix density with the collective matrix
stress tensor. 
 Remembering that the intermolecular forces, ${\bf F}_\alpha(t)$, can
also relax via the dynamics of a tracer segment $\beta$,
 a new projection of $\Gamma_{\alpha\beta}$
leads to:
\beq{16}
\Gamma_{\alpha\beta}(t) \approx  \frac{V}{3k_BT}  \sum_{abc=}^{xyz} 
\int \fraq{d^3k}{(2\pi)^3} \int_0^N d\alpha' d\beta'
V^{ab,\alpha\alpha'}_k
\langle \sigma^{Qb}_{\bf -k}(t) \varrho^{Q\alpha'}_{\bf k}(t) 
\varrho^{Q\beta'}_{-\bf k}  \sigma^{Qc}_{\bf k} \rangle  
V^{ca,\beta'\beta}_k
\eeq
Here, $\sigma^a_k(t)$ is a component of the collective matrix 
stress tensor where one direction is 
longitudinal, i.e. parallel to the wavevector.  
\beq{17}
{\bf \sigma}_{\bf k} = - \sum_{m=1}^\nu \int_0^P d\gamma\; \frac 1k
 ( k_BT {\bf k} - i
{\cal {\bf F}}^m_{\gamma} ) e^{i {\bf k R}^m_\gamma }\; .
\eeq
The equilibrium averages of the components of the stress tensor are connected
to the elastic constants  or high frequency moduli of the liquid
\cite {hansen,boonyip}.
\beq{18}
\langle \sigma^{a}_{-\bf k} \sigma^{b}_{\bf k} \rangle = k_BT V G^{a}_k \delta_{ab}\; .
\eeq
The formal expressions for the microscopic,  elastic constants are known and
connect $G^a$ to the total potential including inter-- and  
intramolecular forces \cite {hansen,boonyip}.
 In order to make a connection to experimental measurements the elastic 
constants not at short, microscopic times but in the glass relaxation regime
would be required. It is the time scale of the glassy or $\alpha$--relaxation
which is connected to the monomeric friction coefficient, $\zeta_0$,
 entering the Rouse model  and our generalization of it \cite{ferry}. 
 Except for simple liquids little theoretical understanding
of the glassy moduli exists \cite{gs,harte}. Since the exact values are not
required in our case, we assume for simplicity:
1) The different elements of the glassy moduli are roughly equal, $G^a\approx 
G$.
2) Their magnitude is $G\approx\varrho_m k_BT$, as would be found in the Rouse
calculation \cite{de}.  With 
these technical assumptions the vertices, which obey
$V^{ab,\alpha\beta}_k=V^{a,\alpha\beta}_k\delta_{ab}$,
simplify to $V^{a,\alpha\beta}_k=V^{\alpha\beta}_k$.

 Non--diagonal
elements in the memory function matrix, $\beta\ne\alpha$, 
arise due to the connectedness of the probe polymer
 and are a central element of the PMC approach. Monomer density
fluctuations, $\varrho_k^\alpha=e^{i{\bf kR}_\alpha(t)}$, 
at different tracer sites are correlated. For a Gaussian ideal polymer the
equilibrium  intramolecular correlations 
are well known \cite{de} and follow from \gl{5}:
\beq{19}
\omega^{\alpha\beta}_k = \langle e^{i {\bf k} ({\bf R}_\beta-{\bf R}_\alpha)} \rangle = 
e^{-\frac{k^2}{6} \langle ({\bf R}_\alpha-{\bf R}_\beta)^2 \rangle } = 
e^{-\frac{k^2\sigma^2}{6} |\alpha-\beta|}\; .
\eeq
The standard normalization in the vertices, see \gl{22} below,
 causes the appearance of the differential operator, 
$(\omega^{\alpha\beta}_k)^{-1}$, the inverse of the ``matrix'' \gl{19}, 
which is defined by:
\beq{20}
\int_0^N d\beta (\omega^{\alpha\beta}_k)^{-1} \omega^{\beta\gamma}_k = \delta(\alpha-\gamma)\; ,\qquad
\mbox{and }\; \partial_\alpha \omega^{\alpha\beta}_k = \pm \fraq{k^2\sigma^2}{6}  
\omega^{\alpha\beta}_k \fur \alpha=0, N \; .
\eeq
The boundary condition simply follows from \gl{4}.
The mathematical form of $(\omega^{\alpha\beta}_k)^{-1}$ can easily be found,
 most simply from looking at the 
continuum limit of the known \cite{kss2,bz3},
 finite matrix inverse $\omega^{-1}$.
\beq{21}
(\omega^{\alpha\beta}_k)^{-1}  = \delta(\alpha-\beta) ( -\fraq{3}{k^2\sigma^2} \partial^2_\alpha
+ \fraq{k^2\sigma^2}{12} ) \; .
\eeq
This leads to the following expressions for the vertices,
where for simplicity the three point correlations are neglected,
i.e. $Q=1$ in \gl{22}.
\beq{22}
V^{ab,\alpha\beta}_k = \int_0^N d\gamma
\langle { F}^{Qa}_\alpha \varrho_{\bf - k}^{\gamma}
\sigma_{\bf k}^{b} \rangle (\omega^{\gamma\beta}_k)^{-1}
\langle \sigma_{\bf -k}^{b} \sigma_{\bf k}^{b} \rangle^{-1}
= - k^a \delta_{ab} \varrho_m h_k \fraq{k_BT}{VG^a_k}
(\omega^{\alpha\beta}_k)^{-1}\; .
\eeq
The approximations concerning the projector, $Q\ne1$, and the values of the
glassy moduli can easily be improved.
Again,
 however, little is known about the dynamics of the four point correlation
function 
$\langle \varrho^\alpha(t) \sigma^a(t) \varrho^\beta \sigma^b \rangle$
in \gl{16}. We  assume that
this matrix of  
friction  functions is dominated by exactly the same dynamical processes which
determine the long 
time tracer diffusion coefficient. Thus, the $\Gamma_{\alpha\beta}(t)$
relaxes (in a parallel fashion) via probe motion described by  
the collective (coherent) single chain correlator evaluated in the
renormalized Rouse model \cite{kss1,kss2}, and the full collective  matrix
correlator, which corresponds to
 the shear modulus,  $G_k(t)$, for our present choice of slow
variables.  In this case in order to account
for the matrix structure 
of $\Gamma_{\alpha\beta}$, the result which adequately
captures the equilibrium, $t=0$, correlations is:
\beq{23}
\int_0^N d\alpha'd\beta'
 (\omega^{\alpha\alpha'}_k)^{-1}
\langle \sigma^{Qa}_{\bf -k}(t) \varrho^{Q\alpha'}_{\bf k}(t) 
\varrho^{Q\beta'}_{-\bf k}  \sigma^{Qb}_{\bf k} \rangle  
 (\omega^{\beta'\beta}_k)^{-1}
\approx  \fraq{V k_BT}{\omega_k} G^a_k(t) \delta_{ab} 
(\omega^{\alpha\beta}_k)^{-1} \omega^{\rm RR}_k(t)\; .
\eeq
Before writing down the results for the friction matrix let us note that $\Gamma_{\alpha\beta}(t)$ 
separates into two independent memory functions because of the structure of $\omega^{-1}$, \gl{21}.
\beq{24}
\Gamma_{\alpha\beta}(t) = \zeta_0 \delta(\alpha-\beta) ( \Sigma'(t) - M(t) \partial_\alpha^2 ) 
\eeq
$\Sigma'(t)$ is another measure of the uniform drag exerted by the matrix on
all tracer segments and is 
closely related to $\Sigma(t)$
of \gl{14}.
 From eqns. (\ref{16}) to (\ref{24}) it follows:
\beq{25}
\Sigma'(t) 
= \fraq{(\sigma\varrho_m k_BT)^2}{36\zeta_0} 
\int \fraq{d^3k}{(2\pi)^3} 
k^4 (\fraq{h_k}{G_k})^2 
G_k(t) \frac{\omega^{\rm RR}_k(t)}{\omega_k}
\eeq
As will become evident from  section 3.B,
 $\Sigma$ and $\Sigma'$ actually agree on intermediate 
time scales. This follows as the integrand in
$\Sigma'$ differs from the one in $\Sigma$, \gl{14}, by a factor
$(\sigma^2k^2/12\omega_k)$, which equals unity for intermediate wavevectors. 
$\Sigma$ and $\Sigma'$  are identical as long as the (major) contributions 
 to the friction of the 
center--of--mass arise from internal modes, i.e. from wavevectors obeying
$kR_g\gg 1$.  They differ somewhat for longer times because, as is evident
from the different boundary conditions, the segment coordinates, ${\bf
R}_\alpha(t)$, and the monomer  
densities, $\varrho^\alpha_k(t)$, cannot be decomposed into the same normal
modes  \cite{de,fractal} for finite $k$.
This  effect of the ends of the tracer chain, however, only affects
 the form of the final relaxation of the memory function.
We prefer $\Sigma(t)$ to
$\Sigma'(t)$ for both simplicity reasons, and the fact that the
well known hydrodynamic long time tail of the diffusion coefficient in
colloidal systems  is recovered
\cite{fractal,hessklein}, 
$\Sigma(t) \to t^{-5/2}$ for $t\to\infty$.
Very importantly, we note that 
 the molecular weight scaling of the diffusion coefficient is {\it not}
affected by this uncertainty about the final power law decay of  the exact 
center--of --mass friction function. 

\begin{figure}[H]
\centerline{\rotate[r]{\epsfysize=18.cm 
\epsffile{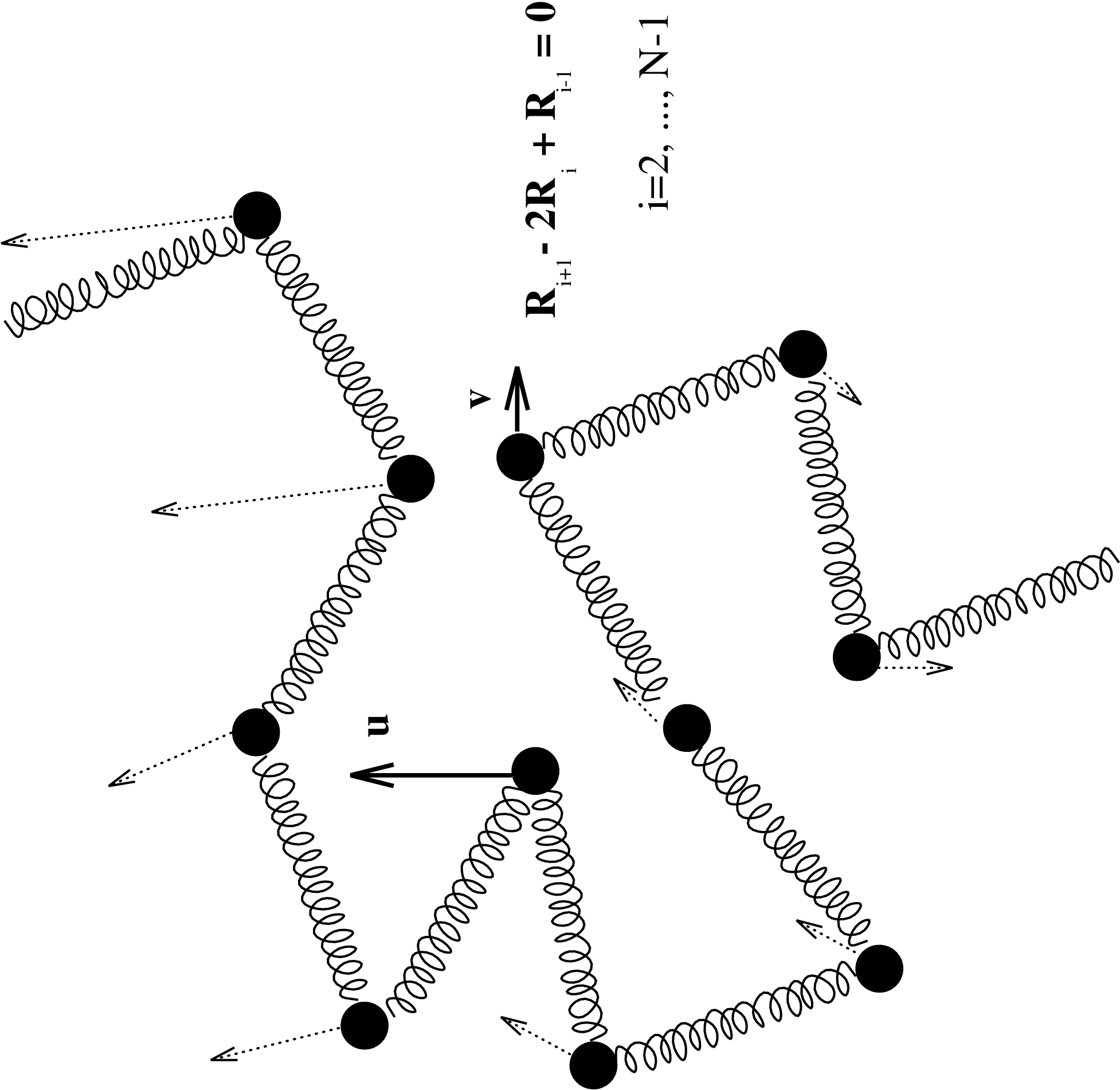}}}
\caption{ }\label{fnmin}
\end{figure}

A conceptually different friction function, $M(t)$, determines the
conformational and single chain stress relaxations \cite{kss2}. 
Eqns. (\ref{16}) to (\ref{24}) lead to:
\beq{26}
M(t)
= \fraq{(\varrho_m k_BT)^2}{\sigma^2\zeta_0} 
\int \fraq{d^3k}{(2\pi)^3} 
 (\fraq{h_k}{G_k})^2 
G_k(t) \frac{\omega^{\rm RR}_k(t)}{\omega_k}\; .
\eeq
In section 3.B the connection of \gl{26} to the previously used expression
\cite{kss2,ks2}  for $M(t)$ is detailed.
Let us discuss the role of the conformational memory function, $M(t)$,
 in the long time, Markovian limit. 
This is possible as the time scale of $M(t)$ is found  \cite{kss2} 
to be shorter than \taud
by a factor proportional to   $1/\sqrt{N}$.
 For simplicity let us also neglect the uniform friction contribution,
 i.e. $\Sigma'(t)$, and consequently study internal, conformational dynamics
 only. 
Eqn. (\ref{1}) then reduces to:
\beq{27}
\zeta_0( 1 - \hat M_0 \partial_\alpha^2 )  \partial_t\; {\bf R}_\alpha(t) - K_S
\partial_\alpha^2\; {\bf R}_\alpha(t) = F^Q_\alpha(t)\; ,
\eeq
where
\beq{28}
\hat M_0 = \int_0^\infty dt\; M(t)\; .
\eeq
For large $\hat M_0$, the disentanglement, or terminal relaxation
 time \taud follows from
\gls{7}{27} and the use of Gaussian intramolecular forces \cite{kss2}: 
\beq{29}
\mtaud = \frac{\zeta_0}{2K_S} \hat M_0 = \frac{\tau_0}{2} \hat M_0\; ,
\eeq
where $\tau_0=\sigma^2\zeta_0/3k_BT$ connects the theoretical with the physical
time. 
Moreover, considering \gl{27} for times short compared to \taud one concludes that the conformational 
dynamics, i.e. the relaxation of the intramolecular distortions,
$\partial^2_\alpha {\bf R}_a(t)$, 
is arrested up to times short compared to \taud. The exact non--Markovian
analysis of refs. 6, 31, and 34 
 rigorizes this statement and identifies the
time to be $\tau^{\rm RR}\approx \tau_0 N^2 \sqrt{N/N_e}$.
Solving the equation
$\partial^2_\alpha {\bf R}_\alpha(t) = {\bf c}_\alpha$ the only time dependence can
enter via the two 
constants of integration. 
Note, that they are not determined by the boundary condition,
\gl{4}, as we are neglecting the center--of--mass motion.  
\beq{30}
{\bf R}^{\rm conf.}_\alpha(t) = \int_0^\alpha d\alpha' ( \int_0^{\alpha'}
d\alpha'' c_{\alpha''} + {\bf u}(t) ) + {\bf v}(t)  
\; .
\eeq
The entanglement process in the PMC theory leads to the prediction of a 
 very cooperative motion of all
tracer segments. 
 Figure \ref{fnmin} shows this in a discrete bead--spring model, where
the standard discretization of 
$\partial_\alpha^2 {\bf R}_\alpha = {\bf R}_{\alpha+1} - 2 {\bf R}_{\alpha} +
{\bf R}_{\alpha-1}$ applies. The restriction  ${\bf R}_{\alpha+1} - 2 {\bf
R}_{\alpha} + {\bf R}_{\alpha-1}=0$, where $\bf c_\alpha=0$ is set for
simplicity, constrains the 
conformational dynamics of the tracer for times shorter than the
disentanglement time, \taud.
From the arbitrary (isotropic)
 displacements, $\bf u$ and $\bf v$, of two randomly chosen segments
the motion of all other monomers is determined.
 The PMC theory generalizes
the reptation ansatz \cite{degennes,de}, that only  
the two end segments can move freely, to a more general, molecular architecture
transcending cooperative motion; e.g. rings and chains are expected to behave
very similarly.
 Further numerical work possibly can find
information about 
the monomer trajectories by studying the arrested magnitudes ${\bf c}_\alpha$.

In summary, in the PMC approach two central 
memory functions  describing dynamical caging of the tagged polymer by its
surroundings  arise in the analysis of entanglements.
Assuming that the projected
 dynamics is dominated by the identical superposition of 
collective intramolecular and collective matrix
structural dynamics, both memory functions are closely
connected.
For linear chains, in contrast to cyclic polymers, a technical complication  
arises from the end monomers but will be neglected as argued above and 
elsewhere \cite{kss2,fractal}. 
The close connection of $\Sigma(t)$ and $M(t)$ immediately leads
to the prediction that from \gls{14}{26}, and (reasonable) results for the
collective matrix dynamics (see below), the asymptotic Stokes--Einstein
ratio obeys \cite{kss2,ks2,fractal}: 
\beq{31}
\eta D / R_g^2 \sim N^0\fur N\to\infty\; .
\eeq
This PMC prediction agrees with the reptation result
 and shows that the
dynamics of entangled polymers differs strongly from  the
Rouse model  \cite{degennes,de}, $\eta^R D^R \sim N^0$. In this paper we will
analyze finite--$N$ corrections to $D$, $\eta$, and \taud
 in order to study whether
experimentally observed deviations from asymptotic scaling laws
 are described by the finite
 size static and/ or dynamic corrections of the PMC theory. 

\subsection{Renormalized Rouse Model}

The collective, single tracer projected
dynamics entering the friction functions,
$\Sigma(t)$ and $M(t)$ \gls{14}{26}, is required. Corrections to the Rouse
model on short time scales have been worked out and discussed in references 
6 and 32.
We will use the Markovian results of this Renormalized Rouse (RR) theory
as we are interested in the transport properties of the PMC model only. 
In the large $N$ limit (effectively frozen matrix), 
the decay of the PMC friction is dominated by  the collective tracer 
dynamic structure factor evaluated in the RR model \cite{kss1,ks2}:
\beq{32}
\omega^{\rm RR}_k(t) = \omega_k \exp{[-\frac{k_BT\, k^2 t}{\omega_k 
\zeta^{\rm RR}} ] }\; ,
\eeq
where $\zeta^{\rm RR}$ is the friction coefficient of the RR model. It results
from a dynamically perturbative calculation of the intermolecular friction and, 
 for large $N$, exceeds the monomeric friction coefficient \cite{ks3}
$\zeta_0$:
\beq{33}
\zeta^{\rm RR} = \zeta_0 (8/27) \varrho_m d^6 g_d^2 \int_0^\infty dk k^2 
\omega_k^2 S_k \to \zeta_0 \sqrt{\frac{N}{N_e}} \fur N \to \infty\; ,
\eeq
where, for
simplicity we drop the regular, molecular weight independent contribution
($\zeta_0$) to $\zeta^{\rm RR}$ and look at times long compared to the Rouse
time, $\tau^{\rm R} = \tau_0 (N/\pi)^2$, where $\zeta^{\rm RR}$ attains its
Markovian value \cite{kss1,ks3}. 
$g_d$ is the value of the intermolecular segment--segment
 pair correlation function,
$g(r)$,  at the excluded volume diameter $d$, i.e. at the distance of
closest approach or contact. The prefactor of the 
 asymptotic limit, $\zeta^{\rm RR} \sim
\sqrt N$, defines the entanglement degree of polymerization, $N_e$ in the
theory. 
It is defined to be the $N$ where in a perturbative, crossover calculation
the increase of the friction due to the slow tracer--matrix interactions
equals the instantaneous friction modeled by $\zeta_0$ in the Rouse picture
\cite{kss4,ks3}.
$N_e$ defines the crossover
 length scale, $b=\sigma\sqrt{N_e}$, and the Rouse mode index,
$p_e=\frac{\pi}{N_e}$, which separates local, Rouse dynamics from the strongly
entanglement 
affected, more collective modes. In the PMC theory the entanglement
effects on local length scales are explicitly removed
 \cite{kss2,ks2}. As a cut--off length or Rouse mode index the quantities
 $b$ or $p_e$ are chosen, respectively. 

For later reference we quote the result for $N_e$ following from \gl{33}
when setting \cite{kss4,ks2,ks3,fractal}
 $\zeta^{\rm RR}(N=N_e)=\zeta_0$:
\beq{34}
N_e = [ \sqrt{3} \pi (16/9) \varrho_m  \sigma^3 \Gamma^{-6} g_d^2 S_0 ]^{-2}
\;, 
\eeq
where $\Gamma=\sigma/d$ is an effective chain aspect ratio \cite{flory,kcur2},
set equal to unity for simplicity throughout this and the following paper.

Without repeating the discussion of the RR model from references
 6 and 32, we recall
that the build--up of friction is evaluated for times short compared to the 
dynamics of the matrix. Thus, the assumption in the RR model 
of a frozen
matrix, $S_k(t)=S_k$, is appropriate for entangled  polymeric
liquids only. Extensions of the theory to the tracer dynamics in 
short, unentangled polymer solutions and melts require a different treatment 
than presented here. The RR results, \gls{32}{33},
can be viewed as a weak coupling and short time limit to
the $\Sigma$--memory function of the PMC description. Namely, the effective
potential or direct correlation function is estimated from the excluded volume
interactions, $c(r) = - V(r) / k_BT = - (4\pi/3)d^3 g_d \delta(r)$, and the
collective dynamics of the tracer is taken from the Rouse model and is assumed
to be rate determining for the fluctuating force relaxation.

\section{Finite Size Effects in PMC Theory}

\subsection{Tracer Shape Fluctuations}

The relaxation of the entanglement constraints and their resulting memory
functions, $\Sigma(t)$ and $M(t)$, progresses via parallel 
dynamical processes or channels. 
If any of these channels were
 ineffective, i.e. its contribution to the friction
did not relax, the entanglements could only relax via the other channels and
 would be slowed down. Besides
the matrix relaxation, which will be discussed in the next section, the
tracer dynamics as described by \gl{32} opens two channels for fluctuating
force relaxation. A
more collective one at small wavevectors is associated with probe
center--of--mass 
translation, and a more local decay
channel at larger wavevectors 
which can be identified with probe shape fluctuations. The first process
relaxes the fluctuating forces via the coherent motion of all tracer segments,
i.e. the center--of--mass dynamics of the
RR model, $\omega^{\rm RR}_k(t)\sim 
N e^{-(k_BT k^2t/N\zeta^{\rm RR})}$ for $kR_g\ll 1$.
Presumably this decay channel is the PMC (isotropic) analog of a coherent
(anisotropic) reptative motion. 
The local probe shape fluctuations for $kR_g\gg 1$ arise from the continuous
spectrum of internal  modes. Eqn. (\ref{32}) approximates those with the well
known first cumulant expression advocated by Akcasu and coworkers \cite{akcasu}
which correctly describes the time--spatial
correlations, i.e. the scaling with  $k^4t$ in the case of the Rouse model
\cite{degennes,de}.

 The shape, or internal mode, fluctuations lead to a speeding up
of the dynamical relaxation of fluctuating (entanglement) forces associated
with dynamically correlated processes on length scales $\ll R_g$. 
It is instructive
to neglect the shape fluctuation mechanism
 and extend the diffusive center--of--mass
dynamics, $\Phi^{\rm RR}_{\rm cm}(t,k) =
  e^{-(k_BTk^2t/N\zeta^{\rm RR})}$,
to arbitrary wavevectors and  compare the predicted behavior
 with the full PMC and the
reptation/ tube results. In order not to vary the wavevector dependent
contributions to the entanglement friction functions this change in the decay
rates has to be accompanied by an appropriate change in the entanglement
amplitudes. Adopting the frozen matrix approximation for simplicity these
adjustments lead, starting from \gls{14}{26},
 to the following memory functions:
\beq{z1}
\tilde{\Sigma}(t)  \propto \frac{1}{\sqrt{N_e}} \int_0^\infty dk\;
k^4 \frac{\omega_k^2}{N} \Phi^{\rm RR}_{\rm cm}(t,k)\; ,
\eeq
\beq{z2}
\tilde{M}(t)  \propto \frac{1}{\sqrt{N_e}} \int_0^\infty dk\;
k^2 \frac{\omega_k^3}{N} \Phi^{\rm RR}_{\rm cm}(t,k)\; .
\eeq
By construction this change does not lead to changes in the asymptotic
scaling of the transport coefficients, but has interesting effects for the short
time asymptote of the disentanglement process.  
Note also that the identical $N$--scaling for the internal dynamics, i.e. 
$\tilde M_0 \sim \hat M_0 \sim N^3$, 
would be obtained even without adjusting the entanglement amplitudes but
only suppressing the shape fluctuation contribution to force relaxation.
 It is only in the center--of--mass
friction function, $\Sigma(t)$ \gl{14}, that neglecting the shape fluctuations
without correcting the amplitudes would lead
 to a non--reptation like Markovian result, $\hat \Sigma_0\sim N^{3/2}$. 

Let us recall from the
reptation/ tube theory that the polymer end--to--end vector
correlation function, $\langle {\bf P}(t)\cdot {\bf P}(0) \rangle$,
 and the shear modulus, $G(t)$, are proportional to the same
function $\psi(t)$, the tube survival function \cite{de}.
 Therefore, they exhibit the
following asymptote in the reptation theory:
\beq{z3}
G''(\omega)/ G_N \propto \langle {\bf P}''(\omega)\cdot {\bf P}(0) \rangle /
R^2_g  \propto  (\omega\tau_0N^3)^{-1/2} \fur \omega\tau_0N^3\gg
1\quad\mbox{reptation }\,  .
\eeq
In PMC theory 
the end--to--end vector correlation function and the shear modulus
are not rigorously
 proportional to each other in general. Elsewhere
\cite{ks1,ks2,fractal}, it was found that in the frequency window
$1/\mtaurr \ll \omega \ll 1/(\tau_0N_e^2)$ the following connections hold:
\beq{z4}
G'(\omega) \sim G_N\;,\qquad\mbox{and }\qquad G''(\omega) \sim G_N
\frac{M''(\omega)}{N}\; ,
\eeq
\beq{z5}
\langle {\bf P}'(\omega)\cdot {\bf P}(0) \rangle \sim R_g^2
\;,\qquad\mbox{and }\qquad 
\langle {\bf P}''(\omega)\cdot {\bf P}(0) \rangle \sim R_g^2
\frac{(\Sigma''(\omega))^{-1/2}}{N^{5/4}}\; ,
\eeq
where the single/ double primes denote  storage/ loss functions. 
The simplified
 diffusive center--of--mass relaxation of the entanglement constraints as
described by the memory functions $\tilde{\Sigma}$ and $\tilde{M}$,
\gls{z1}{z2}, leads to the identical results as the reptation\ tube model, i.e.
\gl{z3}. 
In the frequency window $1/\mtaurr \ll \omega \ll 1/\mtaur$, considering the
reptative--like coherent decay channel only,  then
eqns. (\ref{33},\ref{z1},\ref{z2},\ref{z4},\ref{z5}) immediately 
lead to \gl{z3}. These reptation--like 
results in intermediate frequency windows
have to be contrasted with the correct PMC results \cite{ks1,ks2,fractal}
which, 
in the specified frequency window and using the frozen matrix approximation,
follow  from eqns.  (\ref{14},\ref{26},\ref{32},\ref{z4},\ref{z5}):
\beq{z6}
G''(\omega) \sim G_N (\omega \tau_0 N^{9/2})^{-1/4}\; ,
\eeq
\beq{z7}
\langle {\bf P}''(\omega)\cdot {\bf P}(0) \rangle \sim R_g^2
 (\omega \tau_0 N^{23/6})^{-3/8}\; .
\eeq
Without derivation let us recall the results for higher frequencies but still
below the crossover frequency to the Rouse behavior \cite{ks1,ks2,fractal}:
$G''(\omega) \sim G_N (\omega \tau_0 N^{16/3})^{-3/16}$
and 
$\langle {\bf P}''(\omega)\cdot {\bf P}(0) \rangle \sim R_g^2
 (\omega \tau_0 N^{40/9})^{-9/32}$.

Reptation and PMC theory therefore describe similar friction contributions
acting on the tracer polymer. The asymptotic $N$--scaling of the transport
coefficients   agrees,  but a different mode spectrum or intermediate
time dependence arises from the treatment of the internal shape fluctuations in
the PMC approach.  
Neglecting the shape fluctuations PMC theory
recovers the reptation results as both
theories then assume that entanglements only relax via the coherent
center--of--mass
diffusion of the tracer polymer. Tracer shape fluctuations in the PMC 
equations do not lead to changes in the asymptotic, Markovian results
for the conformational dynamics,
but do affect the 
initial decay in the final disentanglement process. 
It is just in this
intermediate frequency window where the exponents and the
overall shapes of the shear moduli predicted by PMC theory are in much better
agreement with experimental measurements than the reptation/ tube predictions
\cite{ks2,fractal,baumg,jackwint,kannaan}. Note that shallow slopes,
$G''(\omega)\sim \omega^{-0.23}$, are found in the shear moduli of various
polymeric systems \cite{baumg,jackwint,kannaan}, whereas in dielectric
loss \cite{adachi,adachi1,adachi2,adachi3}
often somewhat higher exponents are measured,
$\varepsilon''(\omega)\sim \omega^{-0.21}$ to $\sim \omega^{-0.33}$.

The extensions of reptation including contour fluctuations
lead to shallower slopes in $G''(\omega)$ than pure reptation and also
model the anomalous molecular weight dependence of the shear viscosity
\cite{doicont1,doicont2}.However, in our opinion these approaches do not
accurately reproduce the observed  power law $N$ and $\omega$ scalings. 
 Moreover, recent dielectric tracer experiments of Adachi
and coworkers \cite{adachi,adachi96}
find on the one hand a reptation--like scaling of the
final relaxation time, $\mtaud\sim M^3$,  for large polymer tracers in
entangled polymeric matrices, but in the identical systems a shallow loss
spectrum, $\varepsilon''(\omega)\sim\omega^{-0.21}$ to $\sim \omega^{-0.31}$.
 They conclude this disproves
 the idea that the anomalous
exponents in the viscosity versus molecular weight scaling
and in the dielectric loss
spectrum are connected. Such a conclusion 
is in good agreement with PMC theory, where two very
different physical effects are the source of these behaviors. As discussed
later  in this section, within PMC theory
 tracer shape fluctuations lead to  anomalous frequency power laws for the
initial stages of the disentanglement process, and, 
 finite size effects of the matrix constraints lead to
non--reptative scaling  of the transport coefficients.

\subsection{Matrix Constraint Porosity and Constraint Release}

Contributions to the memory functions $\Sigma(t)$ and $M(t)$, \gls{14}{26},
are characterized by spatially varying or wavevector dependent 
amplitudes and characteristic times.
Within the PMC description one may view the 
static contributions or amplitudes in the mode coupling vertices as the
strengths of the entanglement 
constraints on a length scale $2\pi/k$ on the tracer dynamics.
The time dependence described by the normalized propagators
captures the disentanglement processes. The net 
friction is obtained from the summation of friction amplitudes on all
length scales weighted by the characteristic rates required for decay
of the corresponding constraints \cite{kss2}.
 Let us  first recall the previous results
where the matrix was assumed to be dynamically frozen in the PMC model
\cite{kss2,ks2,fractal}. This corresponds
to setting $S_k(t)=S_k$ in \gl{14}. The relaxation of the entanglement
friction then only 
proceeds via the probe collective dynamics calculated in the RR
model.  In the asymptotic limit of large degree
of polymerization of the tracer the intrinsic ($N$--independent) characteristic
length scales like $b$, $\xi_\rho$, and $\sigma$ become widely separated from
$R_g\propto \sqrt{N}\to\infty$. Thus,
 \gl{14} further simplifies as the wavevector
dependence of the matrix structure factor, $S_k$, and the direct correlation
function, $c_k$, can be neglected. Also, the asymptotic result of \gls{33}{34} 
can be used, and the PRISM Ornstein--Zernicke equation, $S_k = \omega_k +
\varrho_m h_k$ together with \gl{13} leads to the replacement  \cite{kss3}
$S_0 \varrho_m c_0 \approx -1$.
 The suppression of the center--of--mass diffusion coefficient,
$\hat \Sigma_0$ in \gl{15}, then is of the form \cite{kss4,ks2,fractal}:
\beq{35}
\hat \Sigma_0 \propto (g_d \sqrt N)^2\; .
\eeq
The entanglement friction is proportional to the square of the
number of binary segmental  contacts of a pair of interpenetrating
polymer chains 
weighted by a factor proportional to the probability of
contact, i.e. the intermolecular pair 
correlation function at excluded volume distance. In  PMC theory
the fluctuating
forces describing  the entanglement
friction therefore arise from excluded volume interactions of strongly
interacting polymer chains. Their strength is given by the 
 number of two macromolecules pair--contacts, approximately $N^2 / R_g^3$,
and the probability of close contact of two monomers on different
macromolecules, $g_d$. The terminal relaxation time and viscosity follow as
$\mtaud\propto \eta\propto R_g^2/D\propto g_d^2 N^3$.

It is well known from neutron or light scattering experiments that the
collective density fluctuations of a polymeric melt rapidly decay into
equilibrium \cite{richter,patterson,brown}.
 At least within the accuracy of the mentioned experiments,
$S_k(t)$ has decayed to zero at times much shorter than \taud.
 Naively, this contradicts the 
frozen matrix assumption in \gl{14},
which requires that $S_k(t)$ is nonzero on time
scales where the tracer dynamics determines the decay of the entanglement
friction \cite{kss1}.
 However, obviously it is not necessary that the full amplitude
of the collective density fluctuations is frozen in. Strongly molecular weight
dependent contributions to the tracer dynamics are already 
obtained if some small, but
finite, $N$--independent
 amplitude of the matrix dynamics only relaxes  at times of the order
of the probe dynamics as given by \gl{32}. Indeed, this is the experimental
situation for the stress relaxation function $G(t)$, where $G_N\ll \varrho_m
k_BT$. 
 Little is known experimentally
and theoretically about the existence, magnitude and time dependence of an
entanglement plateau in $S_k(t)$ in the melt \cite{richter,patterson,brown}.
In theta solutions, however, the plateau in $S_k(t)$ has been observed
and there the general expectation has been verified that its
amplitude is proportional to the ratio of shear to compressibility
modulus \cite{brown2}. Note that for nearly incompressible 
melts the sensitivity of neutron or light
scattering  is not sufficient for direct observation of
such small amplitudes.

The question of the entanglement amplitude in the slow matrix dynamics
motivated the re-derivation of the PMC
equations projecting onto the collective stress variables in section 2.A.
Experimentally, the entanglement plateau in the stress modulus is well studied
\cite{ferry}.
In this section it will be shown that straightforward  assumptions 
lead to
identical PMC expressions starting from either the choice of
matrix density or matrix
stress fluctuations as a slow collective variable, and that the previous
frozen matrix assumption nicely connects to established concepts.

Let us first 
comment on our treatment of binary polymer--solvent
solutions which we crudely model as an effective one--component polymer fluid.
Adding a small molecule liquid to
a polymeric liquid has two kinds of effects. First, the equilibrium structure
of the polymeric subsystem is changed. The polymer density fluctuations grow
and the osmotic compressibility strongly
increases \cite{degennes,de,flory,kcur2,kcur3}. Locally,
the contact probability of segments on different macromolecules, i.e. $g_d$,
 goes down \cite{kcur2,kcur3}.
 Also the intramolecular structure can be
changed depending on the type of solvent \cite{degennes,flory,kcur3}. 
We will use a simple Gaussian chain model and 
an integral equation approach, PRISM \cite{kcur1,kcur2,kcur3},
 in order to capture these
effects as well as possible and study their consequences on the dynamics of
entangled polymers \cite{ks3,ks4,fractal}.  A second group of effects of adding
a  solvent directly influences the dynamics of the polymer component. 
The local mobility, i.e. $\zeta_0$, is strongly affected \cite{ferry}. The two
component liquid mixture exhibits  further transport processes,
interdiffusion \cite{hansen} or possibly gel--like modes \cite{broch1,broch2},
 which may couple to the polymer density fluctuations.
  Hydrodynamic interactions arise from the instantaneous
but long ranged solvent  motions \cite{degennes,de,zimm}. 
The local mobility affects our overall time scale and is taken from comparison 
with experiments. The remaining two effects described above
are neglected. The standard argument is 
that the interdiffusion is fast following the solvent motion and that
hydrodynamic interactions are screened \cite{de}. Our approach therefore
focuses on the effects of the liquid equilibrium structure on the dynamics but
does not modify
 the basic dynamical equations with varying solvent content and quality.

The long lived  matrix constraints can be written as a wavevector dependent 
amplitude and a normalized time  and (in general)
 wavevector dependent function:
\beq{36}
S_k(t) = S_k f^S_k \Phi_k(t)\;,\qquad\mbox{ where } \; \Phi_k(t\ll\mtaud)=1\; .
\eeq
The normalized correlation function, $\Phi_k(t)$, describes the final 
decay of the matrix constraints which possess the frozen--in amplitudes
$S_k f^S_k$. Faster decay processes, the so--called microscopic and glassy
relaxation processes, are assumed to be completed. They result in a decay
of $S_k(t)$ from its initial value $S_k(t=0)=S_k$ down to the arrested
amplitude, $S_k f^S_k$ where $f^S_k < 1$. The importance of an 
amplitude of the final relaxation step smaller than unity has most clearly
been recognized in the mode coupling theory of the glass transition \cite{gs}. 
There equations for the frozen in amplitudes $f^S_k$ are derived. In
the present context we will describe a physically plausible model and quote
neutron scattering measurements in order to obtain a simple model for $f^S_k$.

The amplitude of the matrix constraints contains two spatially varying or
wavevector dependent factors which describe what might be called 
the ``porosity'' of the entanglement constraints. First, the
equilibrium matrix structure, $S_k$, describes a nonuniform compressibility
of the surrounding polymeric liquid. We assume a simple
Ornstein--Zernicke form which is appropriate for concentrated or (semi--)
dilute theta solutions and for  wavelengths large compared to monomeric
length scales  \cite{degennes,de,kcur2,kcur3}.
\beq{37}
S_k = \frac{S_0}{1+k^2\xi^2_\rho}\; ,
\eeq
where $\xi_\rho$ is the density screening length
or ``mesh-size'', and $S_0=\varrho_mk_BT \kappa_T$, where 
$\kappa_T$ is the isothermal (osmotic) compressibility of the
polymer subsystem.
 The mesh size  can be measured by neutron
or other  scattering techniques \cite{degennes}.
 Alternatively, liquid state theories predict
its magnitude and density dependence in agreement with blob scaling and field
theoretic considerations
 \cite{ks3,kcur2,kcur3}. In PRISM theory  the form of \gl{37} and the
additional result $c_k\approx c_0$ has been derived \cite{thread,threada}.
 Whereas a melt
is described by a density screening length of the magnitude of monomeric length
scales, a less concentrated polymeric  liquid exhibits a much larger
$\xi_\rho$.
$\xi_\rho$ grows with decreasing concentration until, in the dilute solution
limit, it agrees with the radius of gyration of a single polymer chain
\cite{degennes}. In good
solvents the correct form of the density susceptibility nontrivially deviates
from the Ornstein--Zernicke form \cite{degennes} of \gl{37}. 
 As we require $S_k$ in order to
perform wavevector integrals only, we neglect these quantitative
corrections. In 
this extended blob--scaling picture of de Gennes the difference of a good and a
theta solvent is only captured in the different dependence of monomeric length
scales on the actual polymeric density. Whereas in a theta solvent 
the statistical segment 
length scales are density independent, in a good solvent the excluded volume
interaction leads to an effective monomeric size dependent on density
 \cite{degennes,ks3},
$\sigma\sim \varrho_m^{-1/8}$.

In \gl{36}, only an amplitude $f^S_k<1$ of the total density fluctuations is
arrested up to the disentanglement time \taud. We follow Semenov \cite{semenov}
and Genz \cite{genz} in
estimating the magnitude of its spatially homogeneous part, $f^S_0$, as the
ratio of the shear modulus to the bulk modulus, or in our notation 
$f^S_0=S_0/N_e$. Note that in typical
melts where $S_0\approx 0.25$ and $N_e=50$---$300$ one finds 
$f^S_0$ is of the order of $10^{-2}$ or
smaller. 
We shall not repeat the arguments of Semenov and Genz,
but rather derive the same result for
$\Sigma(t)$ looking at the shear stress variables below.
 Clearly, the very local  
density correlations should not be arrested by the entanglement effect
\cite{de,kss4}. We
follow the general conviction, as it is captured for example in the reptation/
tube model or in the PMC theory, that there exists an entanglement length $b$,
$b^2=N_e\sigma^2$, which determines the spatial resolution of the matrix
entanglement constraints. Adapting the results of Ronca \cite{ronca}, de
Gennes \cite{dgplat}  and des Cloiseaux \cite{declois}
about the plateau in the collective single chain dynamic structure factor,
 $f^\omega_k$, to the plateau in the collective matrix structure
factor, $f^S_k$, we arrive at the following simple model:
\beq{38}
f^S_k = ( S_0 / N_e ) \; e^{-(kb/6)^2}\; .
\eeq
The entanglement length in the mentioned
theoretical considerations and in the comparison with neutron scattering 
is defined by $f^\omega_{k=6/b} = e^{-1}$. We have  checked explicitly that the
simplified 
Gaussian shape assumption in \gl{38} does not affect the results of our theory
appreciably.

\begin{figure}[H]
\centerline{\rotate[r]{\epsfysize=18.cm 
\epsffile{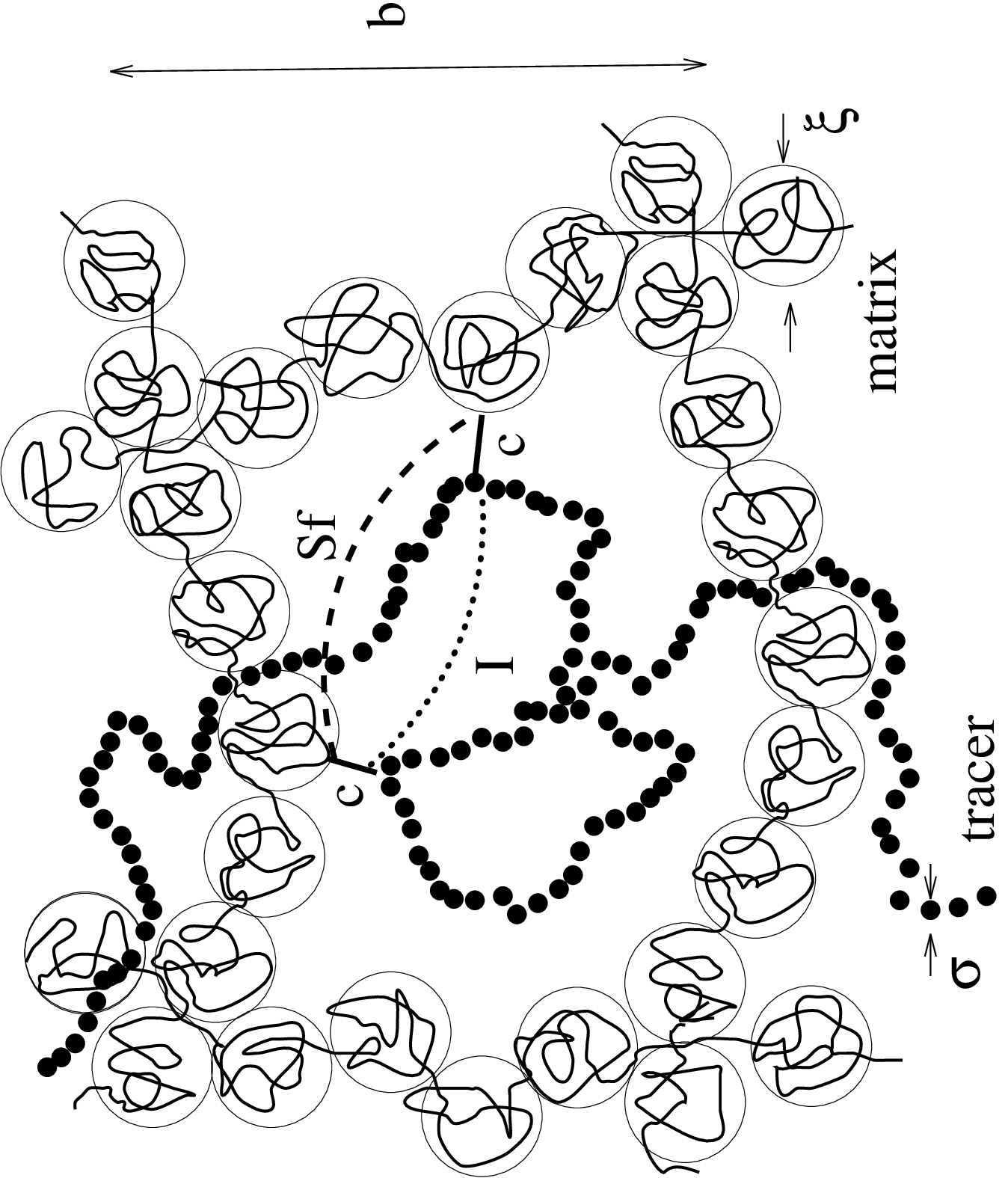}}}
\caption{ }\label{f0}
\end{figure}

The sketch in figure  \ref{f0}
 of the matrix and tracer correlations and characteristic length scales
summarizes the physical factors which enter 
 the PMC memory functions. Clearly, in order to obtain the $N\to \infty$ 
asymptotic
results the tracer size, $R_g$, has to greatly exceed the matrix correlation
lengths, $\xi_\rho$ of the density fluctuations and  $b$ of the elastic mesh.
It will be a central part of our results and their discussion that the
intramolecular correlations of the tracer coarse grain over a much larger
matrix volume in the conformational friction function, $M(t)$,  than in the
center--of--mass memory function, $\Sigma(t)$.

The form of the
 time dependence of the matrix constraints in $S_k(t)$ can be deduced
following the theoretical studies mentioned above
\cite{semenov,genz,dgplat,declois}.
 The matrix provides entanglement constraints up to a time
when the final disentanglement step of the matrix polymers takes place.
The reptation/ tube picture for the collective single chain correlator, and the
PMC results for the generalized Rouse mode correlators \cite{kss2},
indicate that this
disentanglement process is characterized by a uniform, wavevector and Rouse
mode independent
relaxation time \taud. The natural assumption for $\Phi_k(t)$,
 which also agrees
with the theoretical findings of Semenov \cite{semenov} and Genz \cite{genz},
therefore is \cite{kss2}:
\beq{39}
\Phi_k(t) = \Phi(t) = e^{-(t/\mtaud)}\; .
\eeq
Using the eqns. (\ref{36}-\ref{39}) in \gl{14}, and the original expression for
the $M$--memory function  \cite{kss2,ks2,fractal} (containing $S_k(t)$),
 one obtains  the final expressions of
the PMC theory for the tracer diffusion coefficient, the disentanglement time
\taud, and the shear viscosity of linear chain polymers \cite{kss2,ks2}.
 Since the disentanglement time \taud is determined
from the $M$--memory function, \gl{29}, but also enters in  its relaxation,
via \gls{26}{39}, a
self--consistency aspect emerges. Before proceeding to its discussion and
consequences 
for  transport properties of the PMC theory, it is instructive
to derive the same final equations based on the assumption that the collective
stress variables are slow, i.e. starting from \gl{26}.

Eqn. (\ref{26}) was obtained from assuming that the collective matrix stress
variables are the slow variables hindering and slowing down the tracer
dynamics. It is well known that the final stress relaxation is characterized by
the disentanglement time scale \taud so that we can write
\cite{ferry,degennes,de}
with \gl{39}:
\beq{40}
G_k(t) = G_k f^G_k \Phi(t)\; .
\eeq
Neglecting small quantitative differences, the glassy elastic modulus
 can be approximated by:
\beq{41}
G_k \approx \varrho_mk_BT (S_k/S_0) = \frac{\varrho_mk_BT}{1+k^2\xi^2_\rho}\; .
\eeq
Eqn. (\ref{41}) expresses the idea that the glassy, 
elastic stresses are of the order of 
$k_BT$ per segment and their correlation length is the density screening
length; note, that this expression is only needed for inverse wavevectors large
compared to the monomeric sizes.
 It is well established experimentally
that the entanglement plateau in the shear modulus is a factor $1/N_e$ smaller
than the glassy modulus \cite{ferry}.
  We again follow the generally held idea that
the entanglement length $b$ determines the spatial correlations of the elastic
constraints provided by the entanglements. A simple model for the wavevector
dependent plateau in the shear modulus therefore is
\cite{ronca,dgplat,declois}: 
\beq{42}
f^G_k = \frac{1}{N_e} e^{-(kb/6)^2}\; .
\eeq

It does not appear to be a coincidence that the original PMC equations
considering the collective density fluctuations and, the ``new'' PMC equations
resulting from eqns. (\ref{26}) and (\ref{39})---(\ref{42})
which consider the collective stress
variables as slow, lead to identical expressions for the memory functions,
$\Sigma(t)$ and $M(t)$, and consequently to the same theoretical
results. Rather, we surmise that apart from small quantitative differences
identical expressions for the fluctuating force memory functions are obtained
if the same physical assumptions are made. Namely, (1) the fluctuating
intermolecular forces are projected onto the product of a tracer and a
collective dynamical variable. (2) The resulting four point correlation
function is factorized as in \gl{23}, stating that the friction forces relax
via the collective matrix dynamics and the collective tracer density
fluctuations. 
(3) The collective tracer dynamic structure factor is evaluated within the RR
model. (4) There is some long--lived, finite amplitude of the collective matrix
constraints which only relaxes during the final disentanglement process. Its
spatial correlation length is $b$, the entanglement length.  
(5) The disentanglement time follows from self--consistently requiring the
    internal relaxation time to agree with the collective flow time.

Two important new ingredients result from the more detailed treatment of the
matrix constraints as compared to the previous frozen matrix
considerations. First, the constraint release mechanism, i.e. the consideration
of the time dependence of the matrix constraints, relaxes the unphysical
assumption that some collective degrees of freedom do not relax into
equilibrium. Instead, the known slow dynamics of the disentanglement process
enters the theory in a self--consistent way. Moreover, the variation of the
tracer dynamics with matrix molecular weight can now be studied. Second, the
treatment of the constraint porosity is improved. Whereas some previous
numerical studies \cite{ks1,ks2} included the spatial variation of the
compressibility via $\xi_\rho\ne 0$,
 the frozen--in elastic mesh,
characterized by the entanglement length $b$, enters as a new  length
 into the PMC fluctuating force memory functions. An
immediate consequence of $b\ne0$ is the  prediction of
 much larger finite size corrections since the
inequality 
$b\gg\xi_\rho$ holds in general, and finite size corrections within the PMC
theory only
vanish for $R_g$ large compared to all other length scales
\cite{kss2,ks2}.  

Note, that the entanglement degree of polymerization, $N_e$, 
can be estimated from 
the theory, i.e. from the crossover calculation of the RR model,
\gl{34}. However, it can 
also be obtained experimentally from a rheological measurement;
$G_N=\varrho_mk_BT/N_e$. In the
following we will use  the prediction of the present version 
PMC theory that  entanglement corrections enter as a
function of $N$ and $N_e$  in a scaled fashion $N/N_e$;
section 4.B discusses the reasons for this. Thus, we assume $N_e$
to be known experimentally and express results in terms of $N/N_e$.

\section{Asymptotic and General Finite Size Results}

\subsection{Asymptotic Predictions}

The results of the PMC model in the asymptotic, $N\to\infty$, limit have been
worked out previously \cite{kss2,ks2,ks4,fractal}.
We first recall these results for the transport coefficients. 

Changes of the polymer transport coefficients depending on tracer molecular
weight result from the long--ranged contributions ($k\propto R_g^{-1}$)
of the entanglement friction.
The matrix constraint amplitudes and their friction contributions in the
$\Sigma$--
and $M$--memory functions differ due to the different intramolecular weighing
factors, $k^4 \omega_k$ and $k^2 \omega_k^2$ in $\Sigma(t)$ and $M(t)$
respectively. For Gaussian chain polymers a simple but rather accurate
expression for the intramolecular structure factor is \cite{de}:
\beq{omega}
\omega_k = \frac{N}{1+k^2 \, R_g^2 \, / 2}\; .
\eeq
Figure \ref{fintra} shows the intramolecular factors of the entanglement
amplitudes in the $\Sigma$-- and $M$--memory function. Global entanglement
constraints, on the size of the tracer,  dominate  in the
conformational memory function, $M(t)$, only. 
The diffusive tracer dynamics, \gl{32},  enhances  the small wavevector
contributions to the Markovian, transport coefficients.
Figure \ref{fintra} shows that for conformational dynamics and stress
relaxation long wavelength ($kR_g\approx1$) contributions 
extending across the size of the tracer molecule contribute most.
In the center--of--mass friction $\Sigma(t)$  local modes
dominate the amplitude associated with entanglement constraints, although
$kR_g\approx1$ contributions dominate the long time (Markov) friction
coefficient renormalization.
 In  PMC theory,
 this difference in weighting of force correlations is the origin
of the  different, property--specific 
 molecular weight scaling of the corrections to the Rouse,
single polymer dynamics which is so characteristic of entangled polymeric
systems. 

The diffusion coefficient follows from the uniform drag memory function,
$\Sigma(t)$ \gls{14}{15} plus the new analysis of section 3.B:
\beq{43}
 D =  D^R \; [ 1 + \lambda_D \frac{N}{N_e} ]^{-1}\; ,
\eeq
where the Rouse diffusion constant, $D^{\rm R}=(k_BT/\zeta_0N)$, 
is the result for a single macromolecule immersed in a continuum system
characterized by an instantaneous friction coefficient \cite{de,rouse},
 $\zeta_0$.
A reptation like asymptotic scaling is obtained as the corrections to the Rouse
result are of quadratic order in the interaction parameter
\cite{kss4,ks2,fractal} $(g_d\sqrt{N})$.
The strength parameter of the PMC corrections relative to the Rouse result,
i.e. $\lambda_D$, is proportional to a well defined, and in principle
independently measurable, quantity: the mean squared, averaged intermolecular 
force exerted by the matrix polymers on the probe center--of--mass per unit
density:
\beq{44}
\lambda_D = \frac{32}{3\alpha} \sim \langle |F|^2\rangle / \varrho_m\; .
\eeq
It is important to note that in general $\lambda_D$ or $\alpha$, which is
defined  in \gl{44},  are expected to be density and
material dependent parameters. This result differs from the reptation
prediction where $\lambda_D=3$ is a universal number independent of solvent
character, chemical structure, and density of the polymeric system
\cite{de}.

 The inverse
strength parameter $\alpha$ will play the most important role in quantifying
the finite size corrections in the PMC approach,
 and can be more explicitly written as \cite{kss1,ks3}:
\beq{44.5}
\frac{1}{\alpha} = \frac{S_0}{2} ( \varrho_m  g_d c_0 S_0 )^2 
\sim \langle |F|^2\rangle / \varrho_m\; .
\eeq
Note that in previous PMC work, e.g. ref. 33
and earlier publications \cite{kss1,kss2,kss3,kss4,ks1,ks2,fractal},
the literal frozen matrix approximation lead to an expression for $\alpha$ of 
the the form: $1/\alpha\propto g_d^2 N_e$.  As will be discussed in the
following paper no significant
changes in previous results for the concentration dependence
of polymeric transport coefficients arise from this difference
based on PRISM input for simple Gaussian chain models. The actual
magnitudes, however, of the PMC dynamical parameters now come out in almost
quantitative agreement (perhaps fortuitously) with experimental measurements.
The parameter   
$\alpha$ also determines the prefactor of the asymptotic final
disentanglement time, and by assumption, of the shear viscosity:
\beq{45}
\eta = \eta^{\rm R}\, ( 1 + \lambda_\eta (\frac{N}{N_e})^2 \,)\; ,
\eeq
where the Rouse result, $\eta^R=\varrho_m\zeta_0\sigma^2N/36$,  again
follows for a single Gaussian polymer in a continuum with friction coefficient
$\zeta_0$ but without hydrodynamic interactions \cite{de}.
 Here $\lambda_\eta$ is connected to $\alpha$ via:
\beq{46}
\lambda_\eta = \frac{12}{\alpha}\; ,
\eeq
which again can be compared to the reptation prediction \cite{de}
 of $\lambda_\eta=15/4$.
These equations predict a universal ratio of the diffusion
constant  and  shear viscosity in the asymptotic limit:
\beq{47}
\frac{\eta D}{R^2_g} \to G_N \frac{\lambda_\eta}{6\lambda_D} = G_N \frac{3}{16}
 \fur N\to\infty\; .
\eeq
Remarkably, the value of the constant $\lambda_\eta/6\lambda_D=3/16$
is very close to the reptation result of 5/24. Note, however, that the prefactors
$\lambda_D$ and $\lambda_\eta$ are in general
material and thermodynamic state  dependent,
 and that their close connection via the parameter $\alpha$ follows
from two key points of the PMC theory. First, the treatment of the
translation--rotation  coupling discussed in context with \gls{11}{25},
 and second, the assumption
that the entanglement friction in the uniform and the conformational memory
functions, $\Sigma(t)$ and $M(t)$,
 decays via the identical tracer dynamical process, i.e. the RR
collective intramolecular density fluctuations. Of course, the neglect of
hydrodynamic interactions and the simplified treatment of solutions
possibly could also affect \gl{47}.
The exact numerical factors in \gls{44}{46} depend on the exact form of the
intramolecular form factor $\omega_k$ used in the vertices. The values denoted
above result when, for analytical convenience, the simple approximation
\gl{omega} with $R_g^2 = N \sigma^2/6$ is used.

The asymptotic result for the  conformational relaxation time,
\taud, is directly connected with \gl{45}: $\mtaud/\tau_0=N^3/(\alpha N_e)$. 
This result may also be taken as the (experimental) definition of the 
(inverse) entanglement strength parameter $\alpha$.  
Note, that the reptation/ tube result, $\mtaud=3\mtaur (N/N_e)$,
differs from this PMC result in the prediction of an universal value of
the asymptotic prefactor, $\alpha=\pi^2/3$ for reptation, whereas 
 $\alpha$ is given by \gl{44.5} in  PMC theory.
Dielectric relaxation, which measures the end--to--end--vector correlation
function, $\langle {\bf P}(t) \cdot {\bf P}(0) \rangle$,
 does not exactly determine
\taud, as it weights global modes much more heavily
than viscoelastic measurements. From the spectrum of modes obtained
in PMC theory
after neglecting the tracer chain ends as discussed in section 2.A and
refs. 6, 31, and 34,
 one can easily calculate an averaged dielectric
relaxation time, $\tau^\varepsilon=  \int_0^\infty dt \langle {\bf P}(t) \cdot
{\bf P}(0) \rangle / \langle {\bf P}(0) \cdot {\bf P}(0) \rangle$. In a
dielectric 
spectrum, it describes the low frequency behavior,
$\varepsilon''(\omega)\propto \omega \tau^\varepsilon$ for $\omega
\tau^\varepsilon\ll 1$. We find
\beq{dielec1}
\tau^\varepsilon / \mtaur = \frac{\pi^2}{12} [ 1 + \lambda_\varepsilon n ]
\; ,\qquad\mbox{where }\; \lambda_\varepsilon=32/\alpha\; .
\eeq
The difference in the asymptotic limit, $\tau^\varepsilon/\mtaud \to 32/12$,
arises from the contributions of the low lying modes to the
end--to--end--vector fluctuations which are  described by $\Sigma'(t)$ in
\gls{1}{25}. Naturally, the approximation to neglect the chain end effects when
diagonalizing \gl{1} will lead to the largest systematic errors for 
the low lying modes and therefore in quantities like $\tau^\varepsilon$.

\subsection{Parameters and Magnitudes of Finite Size Corrections}

The asymptotic predictions for the transport coefficients of a tracer derived
above
assumed frozen--in constraints of the surrounding polymer
matrix. A simple additivity approximation was used in order to describe the
crossover from  unentangled Rouse to   entangled PMC  dynamics. In the rest of
this manuscript and the following paper, we will discuss corrections to these
results arising from the spatial correlations and the time dependence of the
matrix constraints or entanglements.      The aim is not to describe more
accurately the low molecular weight regime, $N\le 5N_e$, but to study the
effects which prevail up to the highest molecular weights which are 
 experimentally
accessible. As will be discussed, finite $N$ corrections are predicted, which
vanish for $N\to\infty$, but, depending on the situation studied,
 are not yet negligible for $N/N_e\approx10^3$. It is for such corrections 
at  rather high  $N/N_e >5$, that we hope our simple
models for the matrix constraints and the Rouse to entangled dynamics crossover
to be appropriate. As well defined equilibrium structural information is
required, improvements can easily be considered if better (experimental)
information is available. The deviations of the transport coefficients from the
Rouse values depend on molecular weight via the ratios $n=N/N_e$ and $p=P/P_e$
 only, where $N$ ($N_e$) is the tracer and $P$ ($P_e$) 
the matrix polymer  degree of
 polymerization (entanglement degree of polymerization).
 The physical origin of this scaling is the existence of the
 entanglement length, $b$. Entanglements form an elastic mesh of spacing $b$
 and slow down the polymer dynamics only if the intramolecular correlations of
 the tracer extend over distances larger than $b$. The
entanglement friction of a polymer of size $R_g$
therefore depends on the ratio $R_g^2/b^2\sim N/N_e$. 
 This result, however,
rests upon our corrected identification of the frozen--in amplitudes of the
entanglement constraints, i.e $f^S_0 \sim S_0/N_e$ in  \gl{38} or $f^G_0\sim
1/N_e$ in \gl{42}.  

The asymptotic results, \gls{43}{45}, are determined solely
by the rescaled molecular
weight, $n$, and the intermolecular excluded volume strength factor,
$1/\alpha$.   From the comparison with the reptation/ tube predictions for the
diffusion constant and viscosity it appears that a value of $\alpha \approx 3$
should be considered reasonable \cite{de} at least under melt conditions.
 In the following paper 
\cite{pap2} we will give further arguments using PRISM results
and comparisons with
experiments for this choice. The finite 
size corrections  depend on a second, dimensionless
 material dependent parameter, $\delta$,
where:
\beq{57}
\delta = \xi_\rho / b\; .
\eeq
$\delta$
is the ratio of the compressibility length scale (physical mesh),
 $\xi_\rho$, to the elastic
mesh size, $b$. The appearance of this ratio again follows from the spatial
origin of entanglements within the PMC theory. The constraints
arise due to equilibrium,
\gl{37}, and dynamically frozen--in, \gl{42}, intermolecular correlations. 
Again, we generally expect
$\delta$  is a material and density dependent parameter.
 As will
be discussed in the following paper \cite{pap2},
 from the recent comprehensive neutron
scattering and rheological study \cite{fetters},
and theoretical PRISM and computer simulation results,
 values of $\xi_\rho$ and $b$ are known for a
variety of polymers and varying density conditions. Based on this data, 
we estimate $\delta\approx0.05$ for melts and
somewhat larger values for solutions 
\cite{binder}, $\delta\le0.3$.

 PMC theory differs from the
reptation picture  in that it generally predicts a material dependence in the
asymptotic transport coefficients even when expressed in reduced variables.
 Moreover, the appearance of the structural
correlations, parameterized by $\delta$, has no analog in the tube
description, which sets effectively  $\delta=0$. 
This statement does not concern the scaling considerations for the asymptotic
transport coefficients nor plateau moduli  \cite{degennes,degscal,colrub},
but addresses the finite size corrections, i.e. the speeding  up of the
dynamics, due to the correlations of the entanglement constraints.

Introducing the reduced variable, $q=kR_g$, the two memory functions of the
PMC approach,
 $\Sigma(t)$ from \gl{14}, and $M(t)$ from \gl{26}, can be written using
the two non--universal parameters $\alpha$ and $\delta$:
\beq{58}
\Sigma(t) = \frac{2\lambda_D}{\tau_0} \frac{\sqrt{n}}{N^2} 
\int_0^\infty dq \; I^\Sigma(q)\; \hat{S}(q\delta/\sqrt{n})\;
\hat{f}(q/\sqrt{n})  \; 
e^{-[\frac{q^2 (t/\mtaud)}{\Theta\hat{\omega}_q}]} \; e^{-(t/\mtaud)}\; ,
\eeq
\beq{59}
M(t) = \frac{\lambda_\eta}{3\tau_0} \sqrt{n} 
\int_0^\infty dq\;  I^M(q) \; \hat{S}(q\delta/\sqrt{n}) \;  \hat{f}(q/\sqrt{n})
\; e^{-[\frac{q^2 (t/\mtaud)}{\Theta\hat{\omega}_q}]} \; e^{-(t/\mtaud)}\; ,
\eeq
where the reduced compressibility, $\hat{S}=S_k/S_0$, follows from \gl{37}, and
the elastic frozen--in amplitude, $\hat{f}$ normalized to $\hat f(0)=1$,
 from \gl{42}. The first
time dependent factor in \gls{58}{59} arises from the tracer, \gl{32}, the
second from the matrix dynamics, \gl{39}. The amplitudes of the entanglement
forces  are proportional to the spatial weights:
\beq{60}
I^\Sigma(q) = \frac{q^4 \hat{\omega}_q}{\int dq' q'^{2} \hat
\omega_{q'}^2}\; ,
\eeq
\beq{61}
I^M(q) = \frac{q^2 \hat{\omega}_q^2}{\int dq' \hat
\omega_{q'}^3}\; ,
\eeq
where the rescaled intramolecular structure factor, $\hat
\omega_q=\omega_{(k=q/R_g)}/N$, appears. An important quantity specifying the
relevance of the constraint release mechanism compared to the tracer dynamics
is the ratio, denoted as $\Theta$, of the tracer RR time scale  to the matrix
disentanglement  time:
\beq{62}
\Theta(n,p,\alpha,\delta) = \frac{(\zeta^{\rm RR}_{\rm tr}/\zeta_0^{\rm
tr})N^2}{2(\tau_{\rm D}^{\rm m}/\tau_0^{\rm tr})} =
\frac{\tau_0^{\rm tr}}{\tau_0^{\rm m}} \frac{P_e}{N_e}
\frac{\alpha}{2\beta(p,\alpha,\delta)} (\frac
np)^3\frac{\gamma(\delta/\sqrt{n})}{\sqrt{n}}\; ,
\eeq
where $\gamma(\delta/\sqrt{n})=\int dq q^2 \hat \omega^2_q \hat
S(q\delta/\sqrt{n})  / \int dq q^2 \hat
\omega_q^2$ describes the finite size corrections in the RR Markovian
calculation, \gl{33}; $\gamma(x) \to 1$ for $x\to0$.
 Eqn. (\ref{62}) indicates one source of variations
if tracers of chemistry different from the matrix polymers are studied. The
monomeric friction coefficients and the appropriate entanglement molecular
weights will differ in general. However, also the vertex and consequently
$\alpha$, \gl{44.5}, will differ, as the
total intermolecular site--site
correlation function, $h_k$ \gl{13}, is changed by
different tracer--matrix coupling depending on the chemical interactions
\cite{kcur2}. Therefore, we restrict our considerations to chemically identical
tracer and matrix polymers in the following ( except for section 6); i.e. 
$\frac{\tau_0^{\rm tr}}{\tau_0^{\rm m}} \frac{P_e}{N_e} = 1$ in \gl{62}. 

The factor $\beta(p,\alpha,\delta)$ in \gl{62} describes the suppression of the
disentanglement time relative to its asymptotic value:
\beq{63}
\mtaud / \tau_0 = \frac{\beta(p,\alpha,\delta)}{\alpha} \frac{P^3}{P_e}\; ,
\eeq
where $\beta \to 1$ for $p\to\infty$
follows from \gls{29}{59}. $\beta$ is found as the solution of the
self--consistency equation requiring that the matrix disentanglement time
agrees with the longest internal relaxation time of one of the matrix
polymers. 
\beq{64}
\beta(p,\alpha,\delta) = \int_0^\infty dq \; I^M(q) \;
\hat{S}(q\delta/\sqrt{p}) \; \hat{f}(q/\sqrt{p}) \;
\frac{\gamma(\delta/\sqrt{p})}{\Theta(p,p,\alpha,\delta) + q^2/\hat \omega_q} \; .
\eeq
If the matrix disentanglement time, \taud, is found from eqns. (\ref{62}) to
(\ref{64}), the same equations determine the tracer internal relaxation time
with the obvious replacements, especially $\Theta(p,p,\alpha,\delta)\to
\Theta(n,p,\alpha,\delta)$ in \gl{64} and  $p\to n$ elsewhere.  

The shear viscosity including the finite size corrections and the
crossover to the unentangled Rouse result then follows as:
\beq{65}
\eta = \eta^R [\;  1 + \lambda_\eta \; p^2 \; \beta(p,\alpha,\delta) \; ] \; .
\eeq
Note that our crossover model qualitatively, but presumably not quantitatively,
correctly describes the viscosity
at low molecular weights. The longest relaxation time, \taud, however, does not
crossover to the Rouse time, $\tau^R=\tau_0(P/\pi)^2$, with decreasing $P$,
but goes to zero below a certain $P$. It is at such low molecular weights,
$P\approx P_e$, where the self-consistency equation, \gls{62}{64}, loses its
validity. 

From the so--determined matrix disentanglement time, \taud, the diffusion
coefficient of a tracer can be calculated using \gls{15}{58}:
\beq{66}
D = \frac{k_BT}{\zeta_0N} [ \; 1 + \lambda_D \;  n \; 
K(n,\delta,\Theta(n,p,\alpha,\delta))\;  ]^{-1}\; ,
\eeq
where the reduction of the asymptotic entanglement friction constant
 follows from the $\Sigma$ memory function,
\gl{58}:
\beq{67}
K(n,\delta,\Theta) = \int_0^\infty dq \; I^\Sigma(q) \; 
\hat{S}(q\delta/\sqrt{n}) \;  \hat{f}(q/\sqrt{n}) \;   
 \frac{\gamma(\delta/\sqrt{n})}{\Theta(n,p,\alpha,\delta) +
q^2/\hat \omega_q} \; .
\eeq
Again, the crossover to the Rouse result is described qualitatively, but
possibly not quantitatively, correctly. 

In order to compare with  dielectric loss measurements of the
end--to--end--vector fluctuations  \cite{adachi} we must include the finite
size corrections into \gl{dielec1} for the  relaxation time,
$\tau^\varepsilon$, of the end--to--end--vector correlation function of a
tracer in a polymer matrix. Two qualitatively different  finite size
corrections appear in 
$\tau^\varepsilon$ as the conformational memory function, $M(t)$, but also the
homogeneous friction function, $\Sigma'(t)$, affect the low lying, global modes
of an entangled polymer. It is shown in refs. 6, 31, and 34
 that the
generalized Rouse mode correlators of PMC at zero frequency have the following
form: $C''_p(\omega=0)/C_p(t=0)=\tau_0[(\frac{N}{\pi p})^2 (1 +\hat \Sigma'_0)
      +
      \hat M_0]$. One notices the mode independent relaxation rate, $\tau_0\hat
      M_0$ which dominates the viscoelastic dynamics, and the Rouse--like rates,
      which show the familiar $1/p^2$ mode dependence, and are negligible
      except for the lowest modes.
From this, summing over the odd modes  \cite{de}, $p=1,3\ldots$, of the
end--to--end--vector correlation function at zero frequency, one obtains the
dielectric relaxation time,  $\tau^\varepsilon$:
\beq{dielec}
\tau^\varepsilon = \tau_0 \frac{N^2}{12} [\; 1 + (\; 24 \;\beta(n,p,\alpha)\; +
8 \; K'(n,p,\alpha,\delta)\; )\;  \frac n\alpha\;  ]\; ,
\eeq
where $K'$ is the normalized finite size correction arising from the time
integral over the uniform friction function, $\Sigma'(t)$:
\beq{dielec2}
K'(n,p,\alpha,\delta) = 
 \int_0^\infty dq \; I^{\Sigma'}(q) \; 
\hat{S}(q\delta/\sqrt{n}) \;  \hat{f}(q/\sqrt{n}) \;   
 \frac{\gamma(\delta/\sqrt{n})}{\Theta(n,p,\alpha,\delta) +
q^2/\hat \omega_q} \; .
\eeq
$K'$ differs from $K(n,p,\alpha,\delta)$, \gl{67}, in the contributions from the
intramolecular structure only:
\beq{dielec3}
I^{\Sigma'}(q) = \frac{q^6 \hat{\omega}^2_q}{\int dq' q'^{4} \hat
\omega_{q'}^3}\; .
\eeq
The quantity $\beta$ in \gl{dielec} follows from \gls{62}{64} with the replacements denoted
there, and the numerical prefactors in \gl{dielec} assure the correct limiting
behavior, \gl{dielec1}, as $\beta\to1$ and $K'\to1$ for $p\to\infty$.
Again, we suggest \gl{dielec} for intermediate and large
molecular weights of the tracer, $n$, and of the matrix polymers, $p$.
When comparing to experimental data it has to be kept in mind that our present
results apply for chemically identical tracer and matrix polymers
only. System specific variations will arise from differences in the monomeric
friction coefficients, in the molecular weights of entanglement ( see \gl{62})
and from different local packings, i.e. $g_d$, and effective potentials,
$c_0$, in 
\gl{44.5}.

It is one of the central results of the PMC theory that the entanglement
effects on the center--of--mass motion and the conformational dynamics are
closely connected via the two memory functions \cite{kss2,ks2,fractal}
$\Sigma(t)$ and $M(t)$.
They differ in the intramolecular weighting factors,
$I^\Sigma_q/N^2$ \gl{58}, and $I^M_q$ \gl{59}, only. Note from figure
\ref{fintra}  that   $I^M_q$,
\gl{61}, is peaked around $q=kR_g\approx1$, whereas $I^\Sigma_q$, \gl{60},
monotonically increases with increasing $q$.
Friction contributions on the global,
macromolecular size scale determine the internal relaxations. This is one of
the central findings of PMC \cite{kss2},
 which also agrees with the reptation/ tube idea
that a motion of the whole chain is necessary in order to relax internal
fluctuations \cite{degennes,de}. This behavior of $I^M$ 
 strongly differs from the
(tightly connected) finding  that 
contributions to the friction of the center of mass motion, i.e. $I^\Sigma$
\gl{60}, arise from more local correlations \cite{ks2}. See figure
\ref{fintra} where the entanglement amplitudes and the friction contributions,
i.e. the amplitudes weighted by the corresponding relaxation rates (in the
asymptotic, frozen matrix limit), are shown.

This difference causes three major consequences of PMC theory. First, the
asymptotic slowing down of the internal modes is larger by a factor $N^2$ than
the extra friction exerted on
 the center--of--mass \cite{kss2}. Second, the porosity or spatial
correlations of the matrix constraints, $S_k<S_0$ and $f^S_k<f^S_0$, influence
the center--of--mass motion more strongly. The uniform drag friction
function, $\Sigma(t)$, is much more sensitive to the actual values of the
finite, material dependent  length scales, $\xi_\rho$ and $b$. The
conformational friction function, $M(t)$, on the other hand is relatively
independent of these non--universal parameters. Third, the entanglement
friction  can decay via the constraint release mechanism, i.e. the decay of the
matrix constraints following \gl{39}, or via the tracer collective structural
dynamics of the RR model, which exhibits a diffusive pole, \gl{32}. Comparing
the two decay rates for different wavevectors, the tracer dynamics will
dominate the  decay at large wavevectors, $q=kR_g\gg1$, and the matrix
dynamics will dominate 
in the limit $q=kR_g\to0$. Therefore, the constraint release mechanism
will lead to much larger finite size corrections in the conformational
friction function, $M(t)$,  than in $\Sigma(t)$,  as a much larger portion of
the entanglement constraints  arises for $q=kR_g<1$ in $M$ than in $\Sigma$.
The internal dynamics and the viscosities, therefore, will be dominated by the
constraint release mechanism in general, whereas the constraint porosity will
mainly influence the center--of--mass motion. In the following sections we 
discuss these differences and other aspects of the finite size effects in more
detail by numerically
solving eqns. (\ref{58}) --- (\ref{67}) for the physically relevant
parameter ranges, $\alpha\approx 3$ and $\delta=$0 to 0.3.

\section{Predictions for Transport Properties and Relaxation Times}

\subsection{Effects of Constraint Release on the Viscosity}

Considering a melt of polymers with degree of polymerization $N=nN_e$, the
non--linear equation for the finite $N$ suppression, $\beta(n,\alpha,\delta)$
\gl{64}, of the disentanglement time can be easily solved numerically.
 $\beta$ depends on the
molecular weight via the dimensionless ratio $n=N/N_e$ only. The
non--universal 
parameters $\alpha$, where
$1/\alpha\sim\langle|F|^2\rangle/\varrho_m$ in \gl{44.5}, and
$\delta$, \gl{57}, lead to in general 
 density and chemical structure  dependent
results. Results for $\beta$ are shown in figure \ref{f1} spanning a wide range
of possible variations in $\alpha$ and especially $\delta$. 
Since in melts $b\approx$ 35---90\AA\ depending on chemical structure
\cite{fetters}, it
 is apparent from
figure \ref{f1} that the density screening length has to be increased to
unphysically large values, $\delta=\xi_\rho/b\approx0.5$, in order to observe
changes in 
$\beta$ in the range $n\ge5$, where our equations are reliable. From the
discussion in section 4.B of which  spatial correlations dominate the
respective friction functions, it can easily be understood that the
internal relaxation  time \taud is only affected by the constraint release
mechanism but not by the constraint porosity. The disentanglement time, \taud,
and as a consequence the viscosity, $\eta$, follow from the conformational
friction function, $M(t)$, which is most sensitive to constraints for
$q=kR_g\approx1$. These rather long range, intermolecular correlations decay
more effectively via the matrix dynamics than via the diffusive RR tracer
dynamics. The local spatial correlations of the entanglements, either the
equilibrium correlation length, $\xi_\rho$, or the localization length of the
frozen--in amplitude, $b$, influence the conformational dynamics only very
weakly since the local contributions to $M(t)$ are virtually negligible. 

Note that for physically reasonable choices  \cite{fetters,binder} 
of $\delta$, i.e. for
$\delta\le0.3$,
 almost no effects on $\beta$ or \taud can be
seen in figure \ref{f1} for $n\ge 5$. For $n\le 5$  smaller molecular weights
our crossover model cannot be considered very reliable. Thus,
 we conclude that for
experimentally relevant parameter ranges the disentanglement time, \taud, or
viscosity, $\eta$, is suppressed below its asymptotic value by the constraint
release mechanism. To a good approximation the self--consistency equation,
\gls{62}{64}, can therefore be simplified by neglecting the constraint
porosity, i.e. setting $\hat S_q=1$ and $\hat f_q=1$ in \gl{64}. The correction
factor 
$\beta$ then depends on the ratio $n^2/\alpha$ only, i.e. $\beta(n,\alpha) =
\beta(n/\alpha^2)$.
Its asymptotic behavior can be found easily:
\beq{68}
\beta(n/\alpha^2)
\to 1 - \frac 43 (\frac{\alpha^2}{n})^{1/4}\fur n \to \infty\; .
\eeq
The solid line in figure \ref{f1} indicates the solution,
$\beta(n/\alpha^2)$, to the self--consistency equations neglecting constraint
porosity. The asymptotic behavior, \gl{68}, and the independence on $\delta$
for large $n$ can be observed when comparing the solutions,
$\beta(n,\alpha,\delta)$, of \gls{62}{64} including constraint porosity. 
A handy interpolation
approximation to $\beta(n/\alpha^2)$ in the relevant molecular weight
range is given by the following formula:
$\beta(n/\alpha^2)=\exp{-[x+0.22 x^2 + 0.01 x^3 + 0.02 x^4]}$ where
$x=4/3\, (\frac{\alpha^2}{n})^{1/4}$; in figure \ref{f1} it lies on top of the
exact curve, $\beta(n/\alpha^2)$. In the following, i.e. after figure \ref{f2}
in this paper and in the following paper, we will simplify the numerical work
and neglect the constraint porosity corrections in the conformational
dynamics. The above simple but accurate expression for $\beta(n/\alpha^2)$
 will be
used. Errors are thereby made in regions only where the simple matrix and
crossover models we employ are not reliable anyway.

 Figure \ref{f2} shows the predicted viscosities corresponding to
these disentanglement times where the thick lines are given by the simplified
expression neglecting the spatial constraint correlations:
\beq{69}
\eta = \eta^R [ 1 + \lambda_\eta n^2 \beta(n/\alpha^2) ] \; .
\eeq
Again, in figure \ref{f2} it is seen that only unrealistically large values of
$\delta$ lead to appreciable deviations of $\eta$ from the constraint release
result, \gl{69}. Also, even though \gl{69} does not lead to a rigorous 
power law
behavior except in the asymptotic limit, $\eta\sim N^3$ for $N\to\infty$, an
effective power law over two orders of magnitude 
 with an exponent of $\approx$ 3.4 can be accurately fitted to the
results. Note  that the extremely slow approach to the reptation--like
asymptote, $\eta\sim N^3(1-c/n^{1/4})$, where
$c=(4/3)\sqrt{\alpha}$, leads to considerable deviations even
for $n\approx10^3$, molecular weights which are difficult to achieve
experimentally \cite{colby}. In figure \ref{f1}, for values $n=10^3$, the
correction factor, $\beta$, is still of the order 0.7 or less. Experimentally,
of course, the very slow drift of $\beta$ in that range may not be
discernible in the data and an apparent power law with an exponent approaching
3 may be
concluded for $n\ge10^2$---$10^3$. Note that the apparent power law
$\eta\sim N^{3.4}$ holds in a molecular weight range, $n\ge10$, where our
crossover model and the considerations about the matrix structural functions
may be oversimplified, but, are 
qualitatively correct, and contain all the physics
expected to arise in full numerical PMC calculations. As our prediction
 is not a rigorous  power
law in the intermediate $n$ range, it is difficult to estimate what effective
exponents could be reported when fitting power laws to \gl{69}. The results
will also depend on the window in molecular weight of the fits. Exponents up to
3.5 seem easily achievable. 

Two effects of the inverse strength parameter $\alpha$ on the viscosity can be
seen arising from the constraint release mechanism. First, asymptotically the
viscosity becomes proportional to $1/\alpha$. Obviously, the stronger the
intermolecular excluded volume forces are, the higher the viscosities will be in
the entangled regime. Second, increasing $\alpha$ leads to more curvature and
higher effective exponents in $\eta$ as the PMC corrections to the Rouse result
effectively set in at larger $n$ only but then increase more rapidly with $n$.
Finally, since our predicted finite $N$--corrections to the viscosity are
essentially independent of $\delta=\xi_\rho/b$, we predict a sort of
universality of the {\it non}asymptotic behavior nearly independent of both
chemical  structure and polymer concentration when $\eta$ is expressed in terms
of $N/N_e$. This prediction appears to be in excellent agreement with
experiments \cite{ferry,pearson87}. For the values of $n = $ 10---1000 relevant
to experiments, we find that $\Theta$ is typically in the range $<$ 1---0.1.
Thus we conclude that there is no wide separation between the characteristic
single chain conformational relaxation time and the time scale for entaglement
force decay. Hence, a non--Markovian situation is suggested under
experimentally accessible conditions. 

Let us comment on another, non--Markovian consequence of the constraint release
mechanism, which will affect the PMC results for the frequency dependence of
the shear modulus at rather low frequencies, $\omega\mtaud<1$.
Neglecting the matrix dynamics, i.e. the constraint release relaxation,
\gl{39}, and setting $\Phi(t)=1$ in \gl{59}, the conformational friction
function exhibits a long time tail \cite{kss4,ks2,fractal},
$M(t\gg\mtaurr) \to \sqrt{N}\, (\mtaurr/t)^{3/2}$, where $\mtaurr=\tau_o N^2
\sqrt{n}$. This slow decay leads (for finite $N$) 
to an anomalous, low frequency behavior in the
storage part of $M(\omega)$, $M'(\omega) \sim \sqrt{N}\,
(\omega\mtaurr)^{3/2}$, which is more weakly dependent on frequency
than the $M'\sim\omega^2$ expected from a Markovian separation
of time scales. In numerical solutions to the full PMC equations \cite{ksneu},
 neglecting the constraint release, it has been observed that this
anomalous decay of the conformational friction function leads to a final
disentanglement peak in $G''(\omega)$, which is quantitatively broadened for
$\omega\mtaud<1$  compared to the experimental data
\cite{baumg,jackwint,kannaan}. Obviously, the  exponential matrix constraint
relaxation, \gl{39}, yields  $M(t)\propto e^{-(t/\mtaud)}$ (see  \gl{59}) which
will cutoff the long time tail in $M$ at the disentanglement time, \taud. 
Therefore, we expect the quantitative discrepancy \cite{ksneu}
between 
PMC results and experiments for shear moduli 
at low frequencies, $\omega\mtaud<1$, will be resolved 
in future numerical solutions of PMC theory including the constraint release
mechanism. Note, that it also can be expected that no changes will be found for
the high frequency  power law tails of the disentanglement process,
$\omega\mtaud\gg1$,  where the analytic 
PMC prediction  \cite{kss4,ks2,fractal} 
of exponents  around 0.2 --- 0.25 and the
numerical PMC results \cite{ks2,ksneu} agree nicely with shear measurements
\cite{baumg,jackwint,kannaan} but differ strongly with the pure reptation model
prediction of $\omega^{-1/2}$.

\subsection{Constraint Porosity Effect on Diffusion Constants for $P\gg N$}

Upon immersing a tracer polymer with degree of polymerization $N=nN_e$ 
into a polymeric matrix of degree of polymerization $P=pN_e$, the constraint
release mechanism can be neglected if $p\gg n$. Then the matrix polymers are
immobile relative to the tracer. In \gl{62}, the ratio of the corresponding
time scales, $\Theta$,  vanishes as is evident from the factor $(n/p)^3$. The
entanglement friction in the memory functions then can relax only 
via the tracer
collective structure factor (computed with the RR model). The resulting tracer
diffusion coefficients will in general still show finite $n$ corrections as the
matrix constraints are not homogeneously correlated, i.e. $b/R_g$ and
$\xi_\rho/R_g$ are nonzero.
 The tracer center--of--mass experiences the full entanglement
constraints only if it has to distort the frozen--in elastic mesh appreciably
($R_g\gg b$),
and to compress the equilibrium structure uniformly ($R_g\gg \xi_\rho$).
 As the friction
contributions to the conformational dynamics arise from $q=kR_g\approx1$, the
internal conformational 
degrees of freedom are rather insensitive to the spatial finite size
correlations. 
Therefore, in the limit $P\gg N\gg N_e$, the PMC approach
predicts the reptation like
scaling, $\mtaud/\tau_0=N^3/(\alpha N_e)$, for the internal relaxation time of
a tracer polymer.
The center--of--mass motion, however, because the more local
intermolecular correlations  contribute heavily,  \gl{60}, does not feel the
full entanglement constraints as long as the finite length scales, $\xi_\rho$,
$b$ are not negligible relative to $R_g$. The diffusion constants therefore are
increased relative to the $N\to \infty$  asymptotic behavior.
  Moreover, even at fixed scaled degree of polymerization 
$N/N_e$ corresponding to $R_g^2/b^2$, we find that
 the density screening length,
although small, leads to a non--universal reduction of the friction. Note that
the screening length, $\xi_\rho$ (at least in semidilute 
solutions), and the entanglement
length, $b$, are much larger than monomeric sizes, i.e. $\sigma$ or a
persistence length for flexible polymers.
 Therefore, parameterization of the dynamical consequences of these length
 scales  with a few
chemistry dependent parameters, as done in our models \gls{37}{38}, seems
 justified. 

Figure \ref{f3} shows the tracer diffusion constants, $D^{\rm tr}_\infty$, for
two values of the mean square
 force per density parameter, \gl{44.5}, $\alpha=3$
and $\alpha=1/2$, corresponding to $\lambda_D=32/9$ and $\lambda_D=64/3$
respectively. Values of the screening to entanglement length ratio, $\delta=\xi_\rho/b$,
are chosen in a wide range. In the first case of
 $\lambda_D\approx3.6$, which we
suggest to be a melt like case, 
a very rapid crossover from the Rouse, $D\sim 1/N$, to the
asymptotic PMC result, $D\sim1/(\lambda_D N^2)$ from \gl{43}, is seen
for realistic melt values of $\delta\le0.05$, 
Importantly, in the range $n\ge5$ almost no deviation from the
asymptote can be observed. Increasing $\delta$ beyond its physically expected
melt range, somewhat larger exponents in an effective power law, reaching
up to $D\sim N^{-2.3}$ 
may be observed in the intermediate $n$--regime. In the second case,
$\lambda_D\approx 21$, which we consider relevant for solutions
(see the following paper), where also $\delta\le0.3$ is expected,
 an intermediate behavior appears in the range
$1\le n\le50$. There the diffusion constant lies appreciably above its
asymptote. For example, for $\lambda_D=64/3$ and $\delta=0.3$ ($\delta=0.05$)
at  $n=5$ the diffusivity is increased by a factor
$D\lambda_Dn/D^R\approx 3.7$ (1.8), respectively. In
 this $n$--range, a free fit to the numerical results would also lead to
$N$--scaling exponents exceeding the classical reptation result appreciably. 
In figure \ref{f3} a power $D\sim N^{-2.6}$ is drawn for comparison. The
crossover to the true asymptote happens at much
larger molecular weights in the
solution case than in the melt case. 

Note that these deviations from the
reptation like scaling have no analog within the extended reptation/ tube
ideas. Constraint release is irrelevant because of the arrested matrix
dynamics, $P\to\infty$, and the contour length fluctuation mechanism of
 Doi \cite{doicont1,doicont2} does not affect the
$N$--scaling of the tracer diffusion constants.
It appears that a microscopic approach incorporating the liquid structure properly 
is required to
identify these finite size corrections arising from the spatial correlations of
the matrix entanglement constraints. Chemistry and density or composition
dependence is expected for these effects. Also note that in contrast with the
reptation prediction, the asymptotic prefactor of the PMC, $\lambda_D$, is
non--universal and is expected to be polymer--density dependent.

\subsection{Variation of Tracer Diffusivities with Matrix Molecular
Weight} 

The variation of the tracer diffusion constant upon changing the matrix
molecular weight can also be described within our general approach. Because
 of the neglect of the matrix dynamics in the RR calculation the
theory is restricted to the range $p=P/P_e>1$. First, the matrix
disentanglement time has to be found from the self--consistency equation,
\gls{62}{64}. Then the suppression of the entanglement effects on the
center--of--mass motion of the tracer can be calculated from \gls{66}{67}. 
Figure \ref{f4} shows the case with parameters
 argued to be  relevant for polymer melts. 
With increasing $P$ a rather rapid crossover of the tracer diffusion constant
from its Rouse value, $D^R$,
 to the strongly entangled value, $D^{\rm tr}_\infty$, discussed
in section 3.D, is observed. The self diffusion coefficients, $D^S$ for $N=P$,
are also indicated and generally fall in the transition region between the
two asymptotes, $D^R$ and $D^{\rm tr}_\infty$. In figure \ref{f4}, the
parameters $\alpha=3$, $\delta=0.05$ and $\lambda_D=3$, lead to a little 
overshooting of the asymptotic $D^{\rm tr}_\infty$ relative to the
reptation--like  asymptotes. 

The dependence on matrix molecular weight in the
crossover region is not a rigorous
power law but, if approximated by one, corresponds
to exponents of the order of 2. This exponent is smaller than the limiting
ideal behavior of 
$D\sim P^{-3}$
expected from Grassley's constraint release formulation \cite{grassley} and
$D\sim P^{-2.5}$ expected from Klein's analysis \cite{klein1,klein2}. However, if
different material parameters are chosen, in  figure \ref{f5} we use
$\alpha=3$, $\delta=0.3$ and $\lambda_D=18$ corresponding  to dense
solutions, then
larger effective exponents are found. Clearly, the dependence on the
matrix molecular weight must be weaker than the one following from $D^{\rm
tr}\sim  1/\mtaud(P)$, as this result would hold only if the tracer dynamics
was totally frozen in and solely the constraint release mechanism,
or matrix dynamics \gl{39},  would relax
the entanglement constraints in $\Sigma(t)$, \gl{58}.
In this solution--like case a stronger matrix molecular weight dependence in
the transition range is observed compared to the melt case.
$D\sim P^{-2.8}$ is drawn for
comparison in figure \ref{f5}. Again, the self diffusion coefficients lie in
the crossover region.
For the case of figure \ref{f5},
the constraint porosity is also important and leads to deviations of $D^{\rm
tr}$ from the reptation like predictions for all matrix molecular weights as
discussed in section 5.B. Note that the values of the diffusivities in the
solution case lie below the corresponding (same $n=N/N_e$)  melt diffusion
constants because of the larger prefactor, $\lambda_D$, of the large $N$
asymptote. 
 
In figure \ref{5} (and slightly also in
figure \ref{f4}), for the sake of discussion the theoretically predicted
correlation of $\lambda_D$ with $\alpha$, \gl{44}, was violated. In this case
$\alpha$ determines the matrix disentanglement time \taud and the viscosity via
the eqns. (\ref{59}) and (\ref{62}) --- (\ref{65}) but not the strength factor,
$\lambda_D$, in 
the $\Sigma$ memory function, \gl{58},
 describing the tracer. If the matrix viscosities
corresponding to figures \ref{f4} and \ref{f5} were plotted, they would agree,
but the tracer internal relaxation time would differ from the matrix one. A
experimentally relevant situation where such a breaking of the relation in 
\gl{44} can appear is in the 
case of a tracer which differs chemically  from the matrix polymers. Then the
intramolecular correlation function, $h_k$ in \gl{13}, which determines the
vertices or constraint amplitudes, is expected to be different. 
That is, intermolecular tracer--matrix packing is not the same as the pure
matrix--matrix correlations.
Specific
tracer--matrix--polymer chemical interactions caused by the addition of
selective solvent
could also lead to such a more complicated situation.  

\subsection{Variation of Tracer Dielectric Times with Matrix Molecular
Weight} 

The dynamics of a tracer polymer in a polymeric matrix  is slowed
down by the entanglement constraints. 
In the limit of rapidly moving matrix
polymers, the Rouse model with instantaneous friction coefficient, $\zeta_0$,
would apply for the tracer. Dielectric spectroscopy \cite{adachi,adachi96}
measures the slowing down
of the relaxation time of the end--to--end--vector correlation function,
\taue of 
\gl{dielec}. Asymptotically the reptation like scaling, $\mtaue\sim N^3$,
\gl{dielec1}, is predicted by PMC theory.
 
As the global modes contribute to \taue appreciably, in general \taue
differs from the internal, conformational disentanglement time \taud.
 Whenever it is justified to neglect the contributions from the uniform
drag friction, i.e. $c_{K'}=8 K'n/\alpha\ll c_\beta=24\beta n/\alpha$ 
in \gl{dielec}, then the relation
$\mtaue=2\mtaud$ holds. In this case, the conformational friction
determines the dielectric relaxation time, \taue, and therefore
results independent of the spatial constraint correlations, constraint 
porosity, but
strongly dependent on the constraint release mechanism are predicted by PMC
theory. 
Whenever the uniform friction arising from $\Sigma'(t)$, \gl{25}, dominates the
dielectric relaxation time, strong constraint porosity effects resulting from
the finite lengths scales, $b/R_g>0$ and $\xi_\rho/R_g>0$, are predicted in the
PMC approach. Note that the  neglect of chain end effects, discussed in
section 2.A, strongly affects the global mode contributions  in \gl{dielec}.
Without exact numerical diagonalizations of the full equations of motion,
\gl{1}, 
no estimate of the resulting errors in \taue can be given. As figure \ref{f6}
shows this uncertainty mainly affects the results for \taue in the limit of
large tracers in entangled matrices, $n\gg p \gg 1$. Melt--like parameters,
$\alpha=3$ and $\delta=0.05$, are used for figure \ref{f6}.

Our model, \gl{dielec}, rests upon the assumption of strictly
identical matrix and tracer polymers, which may differ in degree of
polymerization only. Due to the approximations in the RR model the results
apply for entangled polymeric matrices, $p=P/P_e>1$ only. Also, the curves are
physically relevant for $\mtaud/\mtaur\approx c_\beta\gg1$ only,
 where the connection between the 
conformational friction function and the final time scale, \gl{29}, holds. Note
that \taud from \gl{63} does not, as would be required physically,
 crossover to the Rouse time in the limit
$N\gg P$ but vanishes asymptotically because of 
$c_\beta\sim p^3/n^{3/2}$ for $n\gg p$. This limitation is not a fundamental
one, but rather a consequence of our simplified modeling of the Rouse to
entangled crossover.

In figure \ref{f6} one notices that apart from small quantitative corrections
the conformational friction determines the dielectric relaxation time, \taue,
for tracers in strongly entangled matrices, i.e. for $n\le p$. For various
matrix molecular weights, 
the dielectric, \taue, and internal, \taud, relaxation
time  follow the asymptotic, $\tau/\mtaur\sim n$, behavior for  $n\ll p$.
For $n>p$, the uniform friction contributions to \taue dominate as 
they increase asymptotically, $c_{K'}\sim p^{3/4} n^{3/8}$ for $n\gg p\gg 1$.
For long tracers in shorter, but still entangled matrix polymers, PMC theory
therefore
predicts an intermediate power law behavior, $\mtaue\sim N^{19/8}$ for $n\gg
p\gg 1$,  resulting from the contributions of the global modes to the
dielectric spectrum. 
The matrix dielectric relaxation times agree with the \taue of a tracer of
identical degree of polymerization, $n=p$, and show a $\mtaud\sim
N^{3.4}$  behavior  for large $n$, as is also observed in the viscosity for
this choice  of  strength parameter,  $\alpha=3$.

The constraint release mechanism and the constraint porosity thus strongly
affect  the molecular weight
scaling of the dielectric relaxation time $\tau^\varepsilon$. Note, however,
that the high frequency wing of the disentanglement process, as described by
PMC theory, is connected to the tracer shape fluctuations. The physical origin
of the fractal frequency behavior, e.g. $\langle {\bf P}''(\omega) \cdot {\bf
P}(0)\rangle \sim \omega^{-3/8}$, \gl{z7}, or  $\langle {\bf P}''(\omega) \cdot
{\bf P}(0)\rangle \sim \omega^{-9/32}$ for higher frequencies, therefore is
predicted to be very different from the one determining the
$\tau^\varepsilon$ scaling with $N$. In the limit $P\gg N$, where reptation
like  scaling in the time scale, $\tau\sim N^3$,  is found since the matrix
constraints cannot decay appreciably, the shape fluctuations of the tracer
still speed up the early stages of the
disentanglement process relative to pure reptation and lead
to a much reduced high frequency slope for $P''(\omega)$ at
intermediate frequencies,
$\omega\mtaud\gg1$; see the discussion in section 3.A.

\section{Extension of PMC Theory 
to Polymer Transport through Random,  Gel--Like Media} 

The diffusion of polymer tracers through amorphous media is of considerable
interest in both physical and biological science
\cite{dullien,muthubaum,andrews}. On the one hand, the original
reptation theory was formulated for the situation of a polymer in a gel
\cite{degrep}. On the other hand, gel electrophoresis constitutes a powerful
but still poorly
 understood technique to characterize polymers, mostly biological
ones \cite{andrews}. In the present section we cannot address the multitude of
different realizations of amorphous materials or of finite driving fields in
electrophoresis. However, the unique aspect of PMC theory is that it connects the
dynamics of polymers to the underlying structure of the surrounding matrix. This
allows one to study  a number of interesting tracer--gel cases within the same
formalism. Moreover, the structural information, i.e. the spatially resolved
compressibility and the mesh of elastic constraints, that uniquely determine
the transport properties of a polymer tracer (if its intramolecular
correlations are specified), raises the possibility of predicting the tracer
dynamics from purely  static experimental data. 

In PMC theory an amorphous solid or gel is characterized by the time independence of
the matrix constraints the probe feels. That is, 
in \gl{36} or \gl{40}, $\Phi(t)=1$.
 Crosslinked polymeric gels  or  silica gels exhibit arrested density
structures which do not relax into  homogeneous, liquid--like
equilibrium. The constraint release mechanism therefore is not operative.

In order to describe gels which are fractal and  exhibit pores on all
(relevant) length scales a simple fractal ansatz for the density correlations
is  used:
\beq{gel1}
S_k = \frac{S_0}{(1+k^2\xi_g^2)^{(D_F/2)}}\; ,
\eeq
where $\xi_g$ is a characteristic mesh or pore size.
Operationally the gel fractal dimension is  defined by a scattering
experiment in the intermediate, self--similar spatial range,
 $k\xi_g\gg1$, where $S_k\sim k^{-D_F}$ is
seen. Exponents in the range
 $D_F\approx5/3$ --- 2 apply to polymeric gels prepared by
crosslinking (good) polymer solutions \cite{degennes}.

 Note that \gl{gel1}
 describes a very simple model of a gel and in reality may not be
adequate for experimental systems. Effects like quenched disorder of crosslinks
\cite{horkay}
 are not described by \gl{gel1}, but can be put into $S_k$  if the
appropriate experimental or theoretical information is available. 
It should be mentioned that quenched crosslinking disorder introduces
heterogenieties \cite{horkay} on an even larger length scale than $\xi_g$  which
results in enhanced values of $S_k$ at small $k$. For example, the following
form has been shown to fit experimental data on some gels \cite{horkay}
\[
S_k = \frac{S_0}{1+k^2\xi_g^2} + S_{xc} e^{-(k\xi_{xc})^{0.7}}\; ,
\]
where $\xi_{xc} > \xi_g$.
In the spirit of our approach, which does not  differentiate between quenched
and annealed disorder, such a structure in $S_k$ would lead to further
nonuniversal finite
size effects in the polymer tracer dynamics extending to even higher tracer
molecular weights as found in the present study.

The density
screening length, $\xi_g$, can be considered to be the size of the largest
pores in the gel. It separates the homogeneous density structure from the
self--similar intermediate range. Of course, the breakdown of \gl{gel1} at
microscopic length scales is neglected for the dynamics of entangled polymers. 

Two distinct
classes of gels can be defined following the discussion of the constraint
porosity in section 3.2. First, in a structurally rigid or ``hard''
 gel, the mass
density structure  can be thought of to pose strong constraints on the motion
of a probe.  The full amplitude of the gel density fluctuations therefore is
expected to contribute to the entanglement friction functions.  
This formally corresponds to:
\beq{gelha}
f^S_k = 1\; ,
\eeq
in \gl{36}. Silica gels presumably belong to this class of hard gels
\cite{ferri}. A second
class includes ``soft''
gels prepared from crosslinked polymer solutions.
Even though the entanglement constraints then are permanent due to the
crosslinking reaction, it cannot be expected that the strength of the so--formed
elastic mesh significantly
exceeds the strength of the non--crosslinked, time--dependent
precursor. The model of section 3.2 should therefore apply and the constraint
amplitudes  again contain a small factor:
\beq{gelso}
f^S_k = (\, S_0 / N_e\,) \; e^{-(kb/6)^2}\; .
\eeq
For lack of detailed knowledge about the spatial correlations of the elastic
mesh in crosslinked gels,
we continue using a Gaussian ansatz for the wavevector dependence of
$f^S_k$. Different $k$--dependent functions  will
affect the comparison with experiments only in so far as shifts in the poorly a
priori known (fit) parameters occur. 

In the soft gel models, the ratio $\xi_g/b$ for crosslinked polymer
gels is expected to be of the order of the values found for
non--crosslinked polymer solutions \cite{degennes}
($\delta=\xi_\rho/b\approx$ 0.3).
  Larger values are expected in the so--called strangulation
regime, where the high crosslinking density forces the entanglement length to
become smaller than the one caused by the temporary crosslinks \cite{lodgegel}.

The question of the intramolecular structure of a polymer in a fractal medium
is very complex and still not well
understood  \cite{muthubaum}. Ideal, Gaussian--like
intramolecular correlations cannot be expected in general. The mass--size
scaling exponent, $\nu$, where $R_g\sim N^\nu$, may even depend on degree of
polymerization, $\nu=\nu(N)$, and on other system--specific features
 \cite{muthubaum}.  We will neglect these
difficulties, and assume a fractal model with fixed mass--size exponent, $\nu$,
for the tracer polymer as well:
\beq{frac1}
\omega_k = \frac{N}{(1+ck^2R_g^2)^{(1/2\nu)}}\; ,
\eeq
where  $c= (1+2\nu)(1+\nu)/d/(2\Gamma(1+1/2\nu))^{2\nu}$. From the obvious
generalization of the $\Sigma$ memory function, \gl{14}, one obtains the
following results for the tracer diffusion coefficient in a fractal gel in
$d$--space dimensions: 
\beq{frac2}
D^{\rm tr} = \frac{k_BT}{\zeta_0N} [\; 1 + \lambda_g \; N^{4-2d\nu}
K_g(R_g/\xi_g, \xi_g/b) \; ]^{-1}\; ,
\eeq
where 
\beq{frac3}
K_g(R_g/\xi_g, \xi_g/b) =
\frac{\int dq q^{d-1} \hat \omega_q^2 \hat S(q\xi_g/R_g) \hat f^S(qb/R_g)\; 
\int dq q^{d-1} \hat \omega_q^2 \hat S(q\xi_g/R_g)}{
(\int dq q^{d-1} \hat \omega_q^2)^2}\; .
\eeq
The dynamics is predicted to depend strongly on the ratio $R_g/\xi_g$.
This dependence of the tracer diffusion coefficient on the ratio of the tracer
size relative to a mesoscopic length scale which characterizes the matrix mesh,
 is closely connected to the constraint porosity effects discussed
 in section 5.2. The latter
have been shown to be important in polymer solutions or if special
tracer--matrix chemical interactions exist.

The strength parameter $\lambda_g$ in \gl{frac2} can,
in principle, be determined from the
tracer--gel structural correlations; i.e. the pair contact value and the direct
correlation function, $c_k$. For example, the PRISM description of binary
polymeric liquids can be evaluated in the limit of vanishing tracer
concentration \cite{kcur2,kcur3}.
 However, because of the possibly varying chemical
interactions of the tracer with the gel pore walls, no general statement about
the magnitude of $\lambda_g$ is possible. Moreover, the increase in the
amplitude $f^S_k$, \gl{gelso}, which likely results from the crosslinking, also
increases $\lambda_g$, but cannot be described theoretically at present.
Note that in the true polymer solution case, i.e. chemically identical tracer
and matrix polymers, \gl{gelso} applies,  $\nu=1/D_F=1/2$, and
$\lambda_g$ is connected to $\lambda_D$,
\gl{44}, via $\lambda_g=\lambda_D/N_e$. For more rigid gels, however, larger
strength parameters, $\lambda_g$, are expected. Moreover, all $\lambda$'s are
proportional to the tracer--gel interaction strengths and therefore are
expected to increase if
the tracer polymers adsorb to the gel pore walls, 
or if other specific tracer--gel or tracer--solvent
interactions exist. Much larger values than
$\lambda_D/N_e$ therefore are possible for $\lambda_g$.

From \gl{frac2} the asymptotic scaling of the tracer diffusion
coefficients with molecular weight can be inferred, as the finite size
correction factor, $K$, \gl{frac3}, approaches unity for large tracers,
$R_g\gg\xi_g$. Asymptotically, $D\sim N^{-5+2d\nu}$ is found
independent of the gel fractal dimension \cite{kss4,fractal}. In
$3$--dimensions for ideal coil tracers the classic $D\sim N^{-2}$ law is recovered.
Note that \gl{frac2}, in the same way as our melt and solution descriptions,
includes a crossover model describing the tracer diffusion constant for all
tracer molecular weights; for the unhindered tracer the Rouse model is again
chosen to apply.
  The PMC results including finite size effects therefore present a
unified description of the tracer dynamics in gels 
arising from the competition of free
and entangled polymer motion, where the second contribution depends on the
length scale ratio $R_g/\xi_g$. 
In this respect the  PMC theory  differs from phenomenological approaches
like the entropic barrier model of Muthukumar and Baumg\"artner 
\cite{muthubaum,baummu1,baummu2,baummu3}, where special geometric
considerations for
$R_g\approx\xi_g$ are invoked,
 and the asymptotic limits ($R_g\gg\xi_g$ or $R_g\ll\xi_g$) and smooth regime
 crossovers
are not included in the
description. 

Figure \ref{f7} shows tracer diffusion coefficients resulting from
\gls{frac2}{frac3} for fractal dimensions corresponding to self avoiding random
walk
polymers, $\nu=1/D_F=3/5$ in the
 Flory approximation \cite{degennes,flory}. Gel
pore 
sizes, $\xi_g$, are denoted by stating the degree of polymerization of a tracer
whose radius of gyration agrees with $\xi_g$, i.e. $\xi_g=R_g(N)$.
Results for two different values for $\lambda_g$ are shown in order to explore
the variation with this unknown system--specific parameter.

 In figure
\ref{f7} one notices that deviations from the Rouse, unconstrained diffusion
start for $R_g/\xi_g$ below 1 but do not lead into the asymptote,
$K_g(x\to\infty,y)\to 1$, up to very high degrees of polymerization
of the tracer. Intermediate effective power law behavior arises
with varying exponents, $D\sim N^{-2.3}$ to $D\sim N^{-2.7}$,
depending on the parameters varied in figure \ref{f7}. 
The steepest molecular weight dependences fall into  a range where the tracer
$R_g$ exceeds the gel pore size $\xi_g$   appreciably.
 This finding appears to disagree with the entropic
trapping ideas 
\cite{muthubaum,baummu1,baummu2,baummu3}, where maximal
non--asymptotic results are
centered around $R_g\approx\xi_g$.
 In the case of PMC theory the finite size effects lead to
anomalous exponents for much larger tracer sizes.

 The results of
\gls{frac2}{frac3} very weakly
 depend on the fractal dimension of the gel. Assuming
for simplicity the identical constraint strengths, i.e. $\lambda_g$ in
\gl{frac2}, the soft gel, \gl{gelso}, and the hard gel, \gl{gelha}, lead to
similar intermediate effective power laws. Somewhat higher exponents result for
soft gels. Increases in the strength parameter, $\lambda_g$, lead to steeper
crossover curves. Note that the existence of the
entanglement length in the flexible mesh systems or soft gels leads to extended
Rouse like behavior as the tracer has to uniformly distort the elastic mesh,
and therefore has to be larger,  in order to feel the full entanglement
constraints.  

The degree of polymerization dependence of the diffusion coefficients, for 
intermediate tracer sizes, cannot be rigorously
 described by power laws. If, however,
power laws are fitted to the numerical results, effective exponents clearly
exceeding the reptation value of 2 are obtained. Note that this holds although the
true asymptotic molecular weight dependence, $D\sim N^{-7/5}$ in figure
\ref{f7}, is
even {\it weaker} than the reptation--like scaling. 
An upper bound to the effective
exponents can be deduced from the small tracer size limit in \gl{frac3}. For
$R_g/\xi_g\to 0$ one easily finds $K_g\sim (R_g/\xi_g)^{2D_F}\sim N^{2\nu
D_F}$, 
which in the limit of a large asymptotic prefactor, $\lambda_g$, and or large
pore sizes leads to the upper bound, $D\sim 
N^{-5+2\nu(d-D_F)}$ where $d$ is the
spatial dimension. In the case of figure \ref{f7} this estimate is $D\sim 
N^{-3.4}$.

\section{Discussion}

The predictions of PMC theory
for the transport properties of entangled polymer melts,
solutions, and gels
 are determined by the intramolecular and intermolecular equilibrium
structural correlations. A number of approximations are necessary in order to
simplify the $N$   generalized Langevin equations  of \gl{1}.

\noindent 1. The fluctuating intermolecular forces are approximated by their
statistical overlap with the collective tracer structure factor and a
collective matrix correlator in accord with known mode--coupling ideas.
 It is expected that for any choice of slow matrix variables 
qualitatively identical PMC expressions result, as shown for the examples of
matrix density and stress fluctuations.

\noindent 2. The slow dynamics of the intermolecular forces is connected to the
diffusive dynamics of the tracer and the (generally nondiffusive) 
 disentanglement process of the
surrounding matrix polymers.

\noindent 3. 
The (projected)
tracer dynamics entering the PMC friction functions is taken from a
 short time and/ or small molecular weight calculation, the RR
model. 

\noindent 4.
The matrix  or constraint release dynamics is evaluated from a
self--consistency argument, requiring that the single polymer and matrix
disentanglement time agree. 

\noindent 5.
Simple models for the strength of the entanglement
constraints exerted by the polymer matrix are formulated. They lead to
qualitatively identical results when considering various slow matrix variables,
as shown for the examples of density and stress fluctuations.

\noindent 6.
In order to calculate bulk transport properties or response functions
 for polymeric liquids, it
is assumed that the entanglement effects arise from incoherently added, single
chain contributions. For example, the bulk viscosities  and dielectric
susceptibility are derived from a single
chain calculation. 

\noindent 7.
Solutions are treated in the same way as melts, neglecting special
dynamical effects, like ``gel--modes'' \cite{broch1,broch2}, but concentrating
on the equilibrium structural changes. 

Assumptions 6. and 7. are familiar from the reptation/ tube theory, and are
generally motivated by the peculiarity of the entanglement effects to
macromolecules (point 6.), and the similarity of these phenomena in melts and
solutions (point 7.).

The robustness of the PMC results with regard to the identification of the slow
matrix variables constraining the tracer polymer (points 1. and 5.) rather
reassuringly verifies that the structural correlations of the matrix
constraints are characterized by two length scales, the entanglement length,
$b$, and the density screening length, $\xi_\rho$. 
Whereas the first length scale is postulated and  identified as the tube
diameter in the
phenomenological  reptation approach, the effects of the second length
scale on the dynamics are neglected there. In PMC theory,
 the matrix constraints are
not fully developed if the ratio $\delta=\xi_\rho/b$ is not small. Smaller
effective entanglement friction and finite size corrections  then arise
and vanish for large molecular weights only. 
The concept of ``lack of full topological correlation of entanglement
constraints'' due to Muthukumar and Baumg\"artner \cite{muthubaum} bears some
resemblance to our ideas, but there are significant differences.

The idea to equate the single chain and collective disentanglement times (point 4.)
appears rather natural and is also suggested by point 6.
In a crude sense it is related to the constraint release mechanisms of 
Grassley \cite{grassley} and
Klein \cite{klein1,klein2} within the phenomenological tube model framework
although there are strong differences. For example, these tube--based
approaches are not self--consistent, and significantly
affect only tracer diffusion and not
conformational relaxation or viscosity in the $N=P$ case.  

Employing the RR collective dynamical structure factor (point 3.) in the PMC
friction functions is suggested by the break down of the Rouse model for
degrees of polymerization above $N_e$. Actually, $N_e$ is defined by
calculating where the Rouse model as a zeroth order approximation is
overwhelmed by the first order correction, the RR model \cite{ks3}. For the
molecular weights and times required in the PMC friction functions, i.e. for
$N\gg N_e$ and $t\gg \mtaur(N_e)$, the RR correlator is physically more
reasonable than the Rouse correlator. Nevertheless, this aspect of the theory
seems the most ``uncontrolled''.
 The use of a fully self--consistent
correlator, i.e. entering the PMC correlator into the friction functions of
PMC theory, is not 
appropriate, 
if naively implemented, since it results in
 a severe  overestimation of the entanglement friction and
an arrest 
(``macromolecular scale localization transition'')
at a finite degree of polymerization \cite{kss4,ks2}.

The determination of the decay of the intermolecular forces from the
collective dynamic structure factor of the tracer bears close connection to one
of the central assumptions of reptation: the coherent center--of--mass 
motion of
the chain in the tube determines the conformational relaxation
\cite{degennes,de}. In agreement with this picture, PMC theory
finds, when neglecting
tracer shape fluctuations, identical results for the conformational dynamics as
reptation. Not only the $N$--scaling of the
internal relaxation time and viscosity agree, but also
the shear stress spectrum shows the high frequency asymptote,
$G''(\omega\mtaud\gg1)\sim \omega^{-1/2}$, familiar from the tube survival
function of reptation \cite{degennes,de}. In better agreement with measurements
of shear and dielectric spectra, PMC results \cite{kss3,ks2,fractal}
including tracer shape fluctuations find
more shallow slopes, $G''(\omega\mtaud\gg1)\sim \omega^{-x}$, where $x\approx$
0.2 --- 0.25. In agreement with recent dielectric
measurements of Adachi and coworkers \cite{adachi,adachi96}, these
non--reptation like power law behaviors are predicted to persist even, if,
 for
strongly entangled polymer matrices, the reptation like scaling of the internal
relaxation time, $\mtaud\sim N^3$, is observed.  

Whereas the tracer shape fluctuations only influence the initial decay of the
disentanglement process, there arise two finite size effects in PMC theory which
affect the transport coefficients of entangled polymers. Both are closely
connected as they arise from the consistent model for the matrix entanglements
discussed in section 3.B. First, a decrease of the entanglement friction
results from the time--dependent
decay of the matrix constraints, termed constraint release
mechanism. Clearly, this idea bears similarity to Grassley's \cite{grassley} or
Klein's \cite{klein1,klein2} ideas for tracer diffusion (but not viscosity),
 although the actual implementations 
strongly
differ. PMC theory describes the decrease of the effective friction coefficient
whereas in refs. 15---17
 two independent relaxation rates
(or diffusion constants) 
are added based on the assumed statistical independence of the reptation and
constraint release transport processes.
 Second, the spatial correlations of the matrix constraints, termed
constraint porosity, enhance the tracer motion as the full constraints are
effective only if the intermolecular forces are summed over finite spatial
regions. The constraint correlations are of two distinct origins:
spatial compressibility,
characterized by the density screening length or mesh size, $\xi_\rho$, and elastic
mesh or entanglement length, $b$, correlations. 

\begin{figure}[H]
\centerline{\rotate[r]{\epsfysize=18.cm 
\epsffile{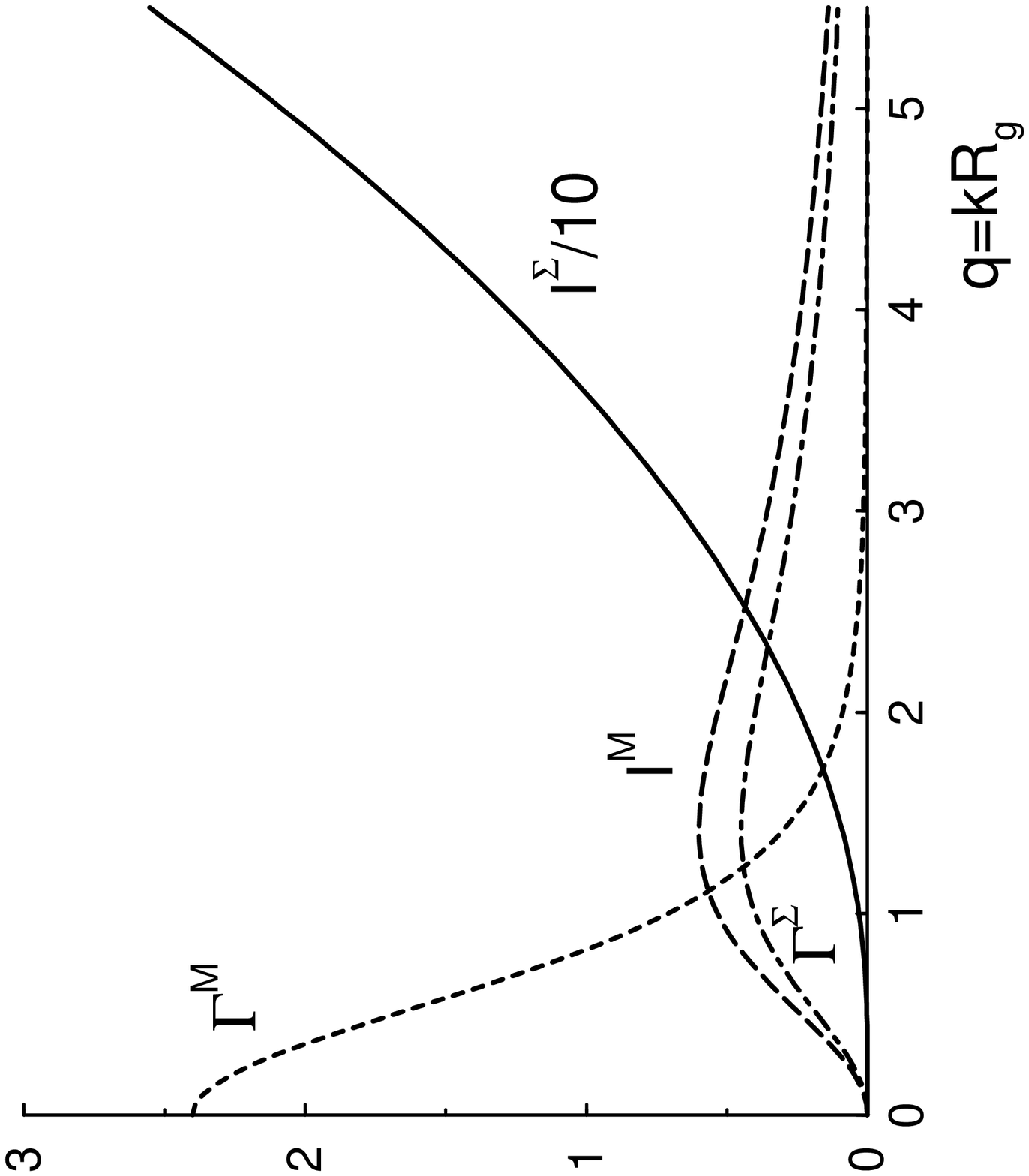}}}
\caption{ }\label{fintra}
\end{figure}

One of the central findings of the PMC approach
is the result that entanglement constraint
 contributions to
the conformational dynamics arise from globally, across the tracer, correlated
intermolecular forces, whereas the center--of--mass friction
results from more local intermolecular forces \cite{kss2}.
The corresponding intramolecular factors determining the constraint amplitudes
are either  peaked around $kR_g=1$ or increase monotonically, see figure
\ref{fintra}.  This finding appears connected to the reptation/ tube idea that
the mobility of the polymer in the tube can be motivated by the picture of
``pulling a wet rope through a tube'' \cite{degennes}. The motion of the
whole ``rope'' is slowed down by short ranged fricition.
Whereas, conformational
relaxation requires  the 
decay of the tube correlations due to the motion of the chain ends
\cite{degennes,de}. Also, the aspect of an underlying diffusive collective 
tracer dynamics, i.e. the collective RR dynamical structure factor, agrees with
this picture. 

A first consequence of the difference in the spatial correlations of the
conformational, $M(t)$, and the uniform, $\Sigma(t)$, memory functions is their
very different Markovian value\cite{kss2},
 $\hat M_0\sim N^3$ and $\hat \Sigma_0\sim N$.
 Because of this feature of  strongly different effects of the
entanglements on the internal and the center--of--mass friction,
 the prediction of conformational and stress relaxation arrest (plateau)
but continued (but slowed down anomalous) segmental diffusion follows.

A second consequence, worked out in this paper, is the very different
sensitivity to finite size effects predicted by PMC theory for internal and
center--of--mass dynamics. Again, because of the rather long ranged spatial
correlations of the entanglement friction in the conformational dynamics,
constraint porosity is irrelevant there. However, the constraint release
mechanism strongly speeds up the conformational dynamics as it overwhelms the
slow tracer RR dynamics on long length scales or for small wavevectors. The
center--of--mass motion, on the other hand, is accelerated due to the decrease
of the entanglement friction on local, finite length scales. The constraint
porosity, therefore, enhances the diffusion constants of tracer polymers even
in matrices of effectively immobile matrix polymers.

Two nonuniversal, microscopically defined
 parameters, the inverse entanglement strength parameter,
$\alpha$, and the length scale ratio, $\delta=\xi_\rho/b$,
 are predicted to control the finite size corrections to the ratios of the 
transport coefficients relative to their Rouse values if
molecular weights are normalized by the corresponding
entanglement molecular weight. The
first parameter exists in the reptation/ tube approach, where an universal
value, $\alpha\approx$ 3--4 is calculated. The second parameter has no
analog there. 

PMC theory
predicts an extremely slow approach of the disentanglement time, \taud,
and hence the viscosity, to their asymptotic, $\eta\sim\mtaud\sim N^3$,
behavior. Higher exponents result from fits of effective power laws to the
numerical PMC results  for  molecular weights $n=M/M_e<10^3$. 
No clear separation of the final disentanglement time from the time scale for
entanglement force decay is expected in experimentally relevant parameter
ranges. 

Diffusion constants for polymer tracers in matrices of immobile polymers are
unaffected by the constraint release process but differ from the asymptotic
behavior, $D\sim N^{-2}$, because of the spatial variation of the entanglement
constraint amplitudes. Figure \ref{f3} exemplifies that for the experimentally
relevant  intermediate
molecular weights effective exponents, $D\sim N^{-x}$, with $x$ 
significantly exceeding the
classical value of 2 will often occur.

The variation of the tracer diffusion constants upon varying the molecular
weight of the matrix polymers results from a combination of both types of
finite size
effects. Strong variations are obtained due to the constraint release mechanism
for systems where the matrix polymer molecular weight is smaller or slightly
larger than the tracer molecular weight.

Constraint release and porosity also compete in the finite size corrections of
the end--to--end--vector relaxation time, \taue, measured in dielectric
spectroscopy \cite{adachi,adachi96}. In the limit of immobile matrix polymers,
the
disentanglement and dielectric times are proportional and follow
reptation--like 
scaling, $\mtaud\sim\mtaue\sim N^3$. In the limit of large tracer polymers in
entangled matrices, a new behavior of \taue is predicted, $\mtaue\sim
N^{19/8}$. It results from the global mode contributions in the dielectric
spectrum and requires $M \gg P\gg M_e$, where $M$, $P$ and $M_e$ are the
tracer, matrix polymer and entanglement molecular weight, respectively. 

The dynamics of tracer polymers in amorphous solids, especially in gels, does
not require qualitatively
 new physical effects to be appended to the PMC description once the physically
 appropriate
 generalizations of the matrix and tracer structure are  made. Obviously, in an
 amorphous solid the 
entanglement constraints cannot fully relax, and hence the constraint
 release mechanism  is absent. The constraint porosity, however, leads
 to strongly molecular weight dependent tracer diffusivities, especially, as
 the 
 tracer--gel interactions may often lead to stronger intermolecular forces
 than in  simple homopolymer systems. For non--Gaussian intramolecular
 correlations of the tracer polymer, $R_g\sim N^\nu$, PMC theory 
does not predict
 reptation like results, but rather
 $D\sim N^{-5+2d\nu}$ asymptotically in $d$ spatial dimensions. 
Very much stronger intermediate molecular weight dependences are predicted,
however, arising from the constraint porosity corrections as long as the ratio
of gel pore size, $\xi_g$,  to tracer size, $R_g$, is not negligible. In
the limit of large tracer--gel interactions an upper bound, $D\sim
N^{-5+2\nu(d-D_F)}$, may be approached, where $D_F$ is the fractal dimension of
the gel. These stronger effective molecular weight dependences do not indicate
in any obvious manner the existence of qualitatively different
 mechanisms of polymer transport, but emerge as natural
generalizations of the constraint porosity effects predicted by PMC theory
also for  
polymer melts and solutions. Indeed, this is a very important point applicable to
all our new results. Traditionally, a change in $N$--scaling exponents is
interpreted in terms of a new dominant transport mechanism at the level of
individual polymer trajectories. In contrast, PMC theory focuses on the entanglement
friction, and the diversity of possible scaling laws emerge as a consequence
of the influence of structure and multiple competing length scales and
relaxation channels on fluctuating force time correlations. Real space physical
motions do not need to be guessed as in phenomenological theories, nor can we
unambiguously infer them.

In the following paper these theoretical predictions will be tested by
quantitative
comparisons with experimental data. 
A necessary preliminary is the estimation of  the equilibrium parameters
specifying the 
dynamics which will be accomplished using integral equation theories,
computer simulation results,  and
experimental scattering and rheological data.

\bigskip
\bigskip

\leftline{\large{\bf Acknowledgments}}
\bigskip
\noindent 
Partial financial support by the Deutsche Forschungsgemeinschaft under grant Fu
309/1-1, and the United States National Science Foundation MRSEC program via
grant number  NSF-DMR-89-20538,  are gratefully acknowledged.

\begin{figure}[H]
\centerline{\rotate[r]{\epsfysize=18.cm
\epsffile{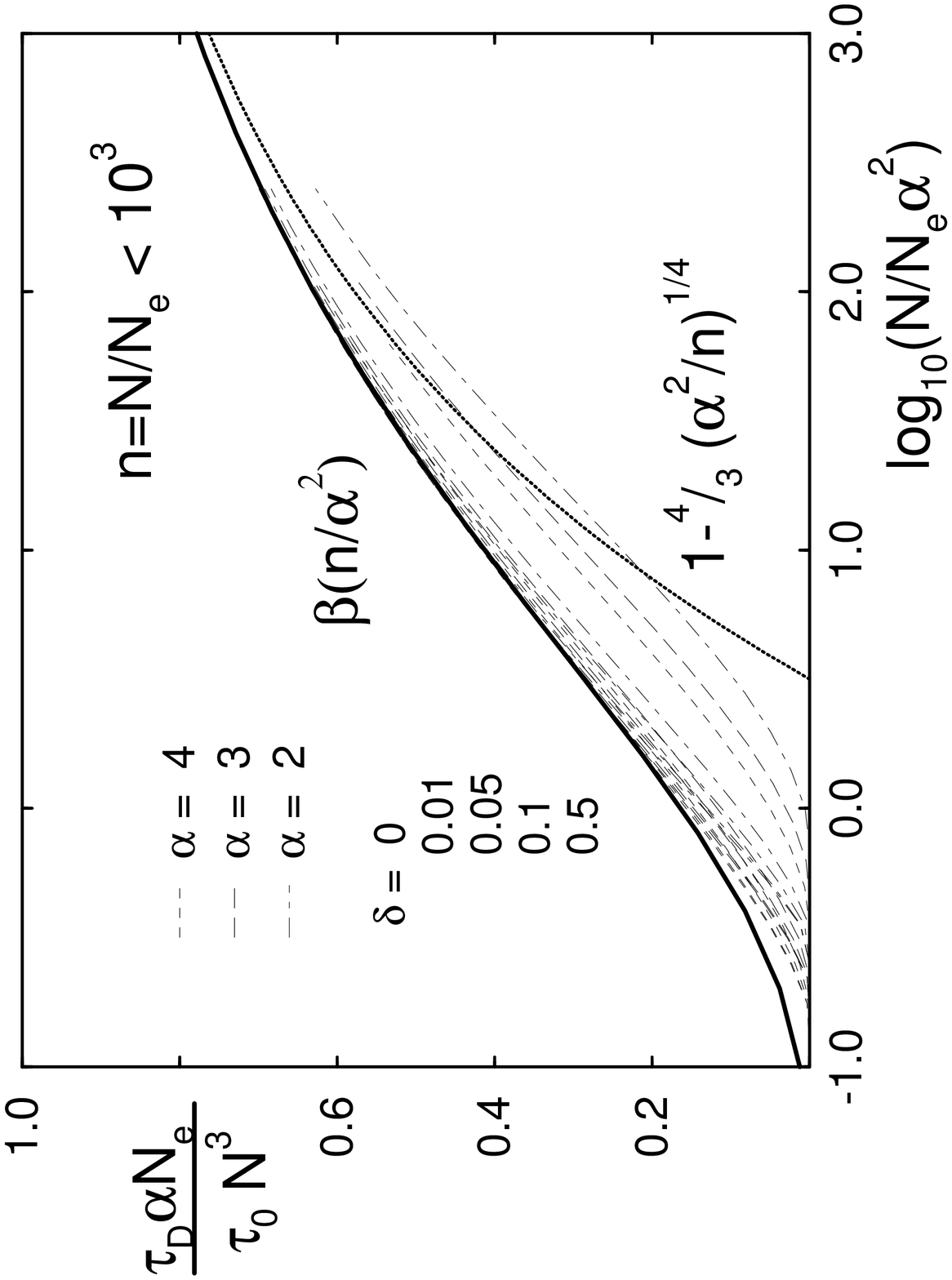}}}
\caption{ }\label{f1}
\end{figure}

\begin{figure}[H]
\centerline{\rotate[r]{\epsfysize=18.cm 
\epsffile{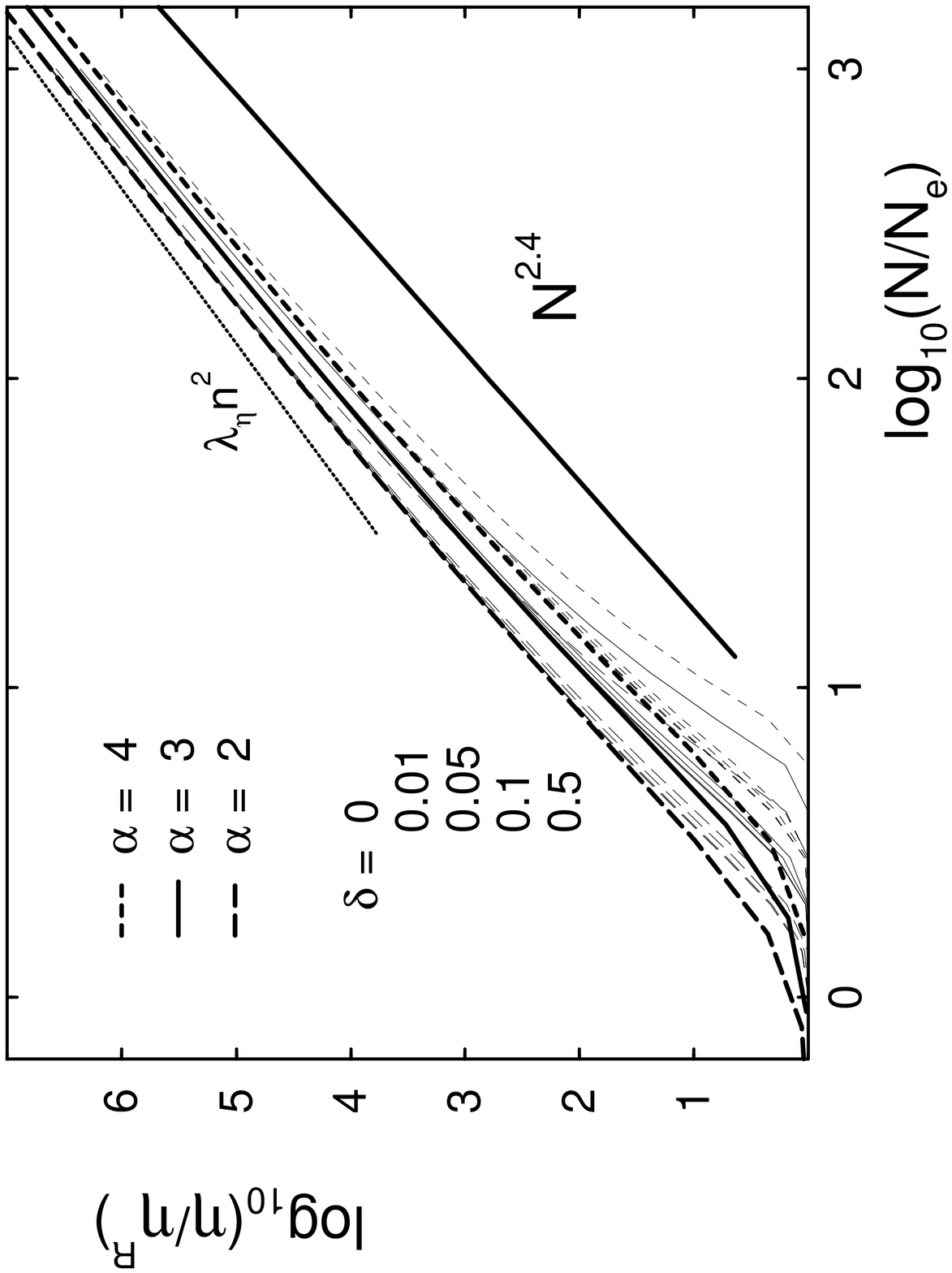}}}
\caption{ }\label{f2}
\end{figure}

\begin{figure}[H]
\centerline{\rotate[r]{\epsfysize=18.cm 
\epsffile{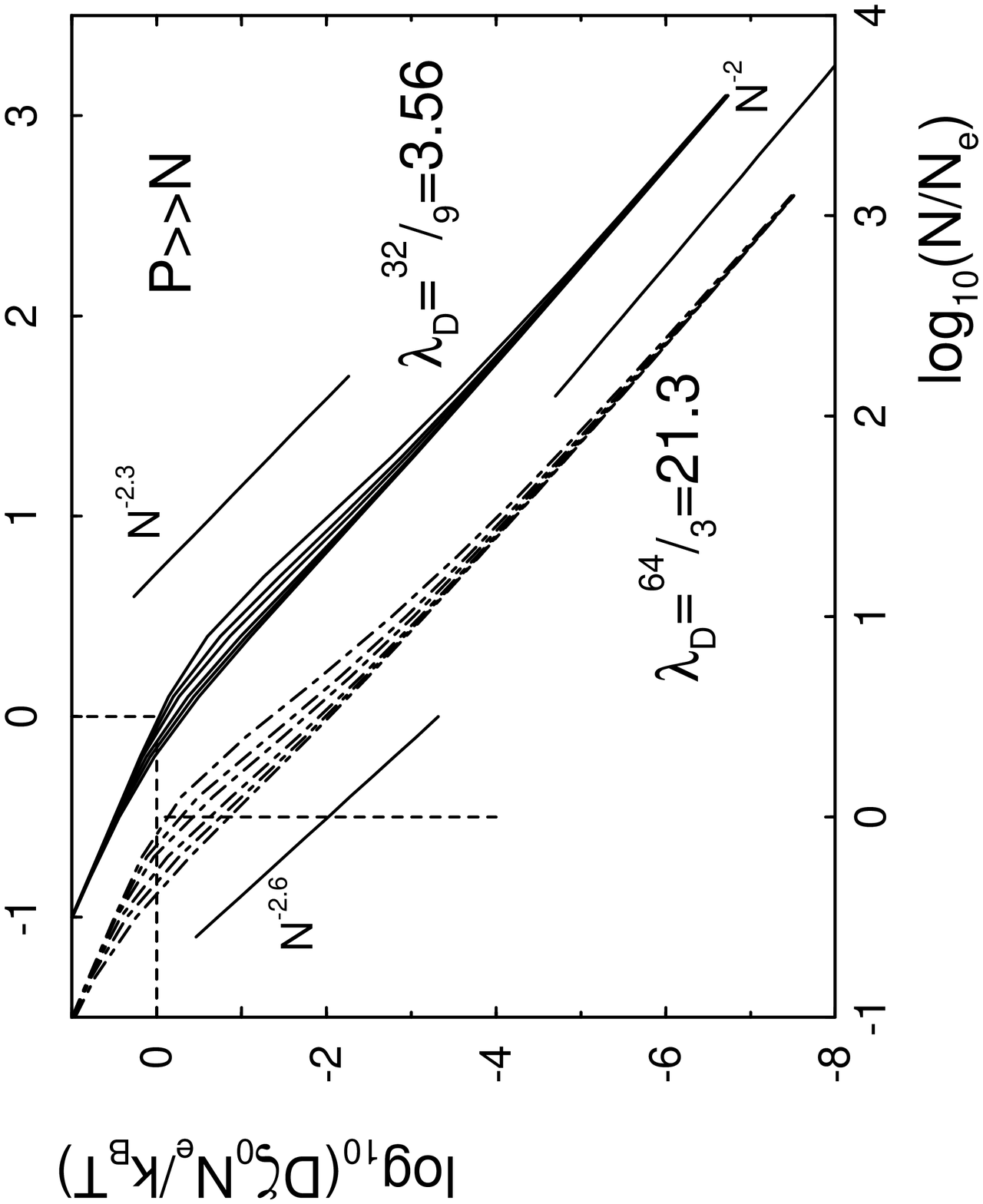}}}
\caption{ }\label{f3}
\end{figure}

\begin{figure}[H]
\centerline{\rotate[r]{\epsfysize=18.cm 
\epsffile{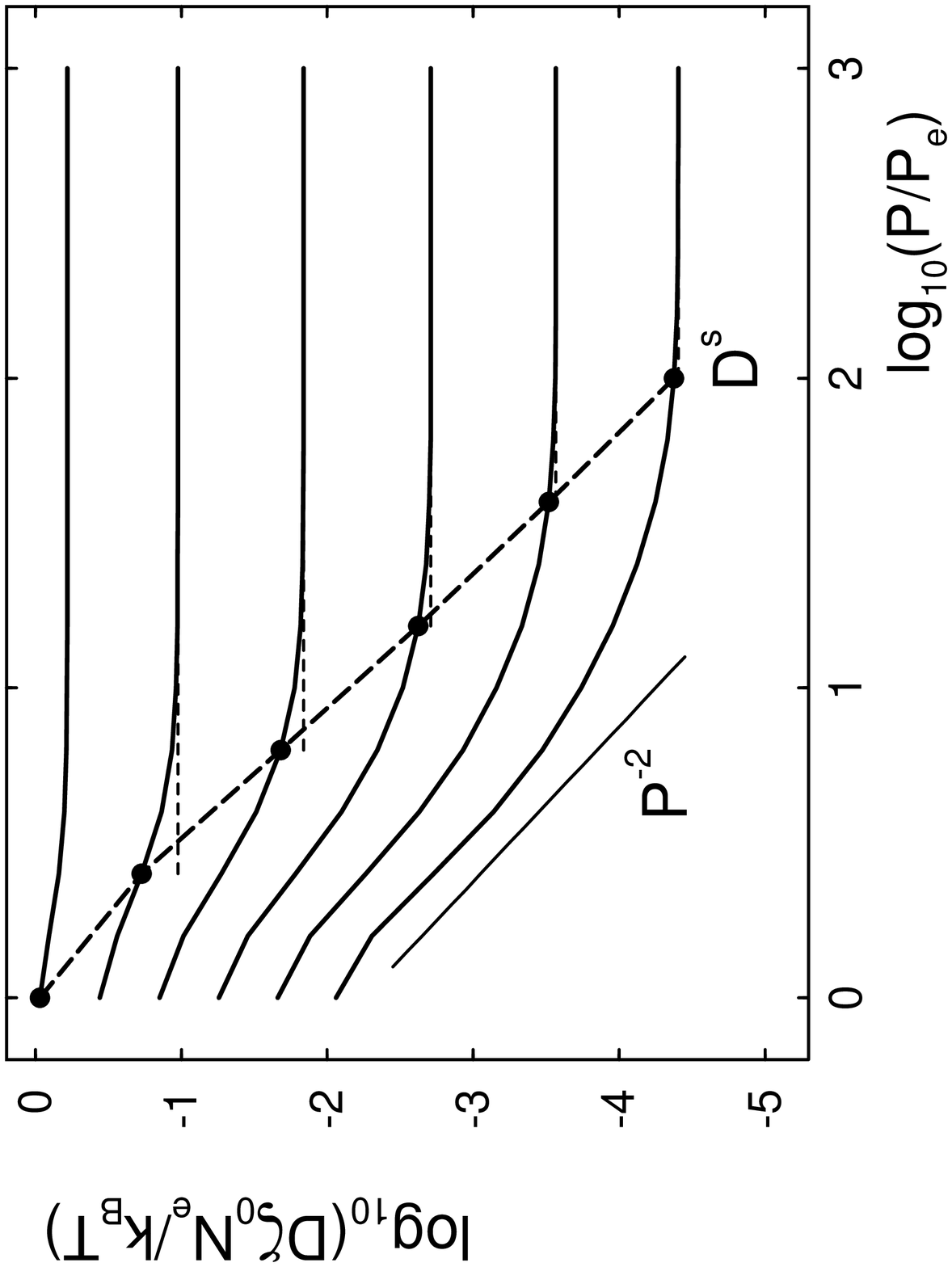}}}
\caption{ }\label{f4}
\end{figure}

\begin{figure}[H]
\centerline{\rotate[r]{\epsfysize=18.cm 
\epsffile{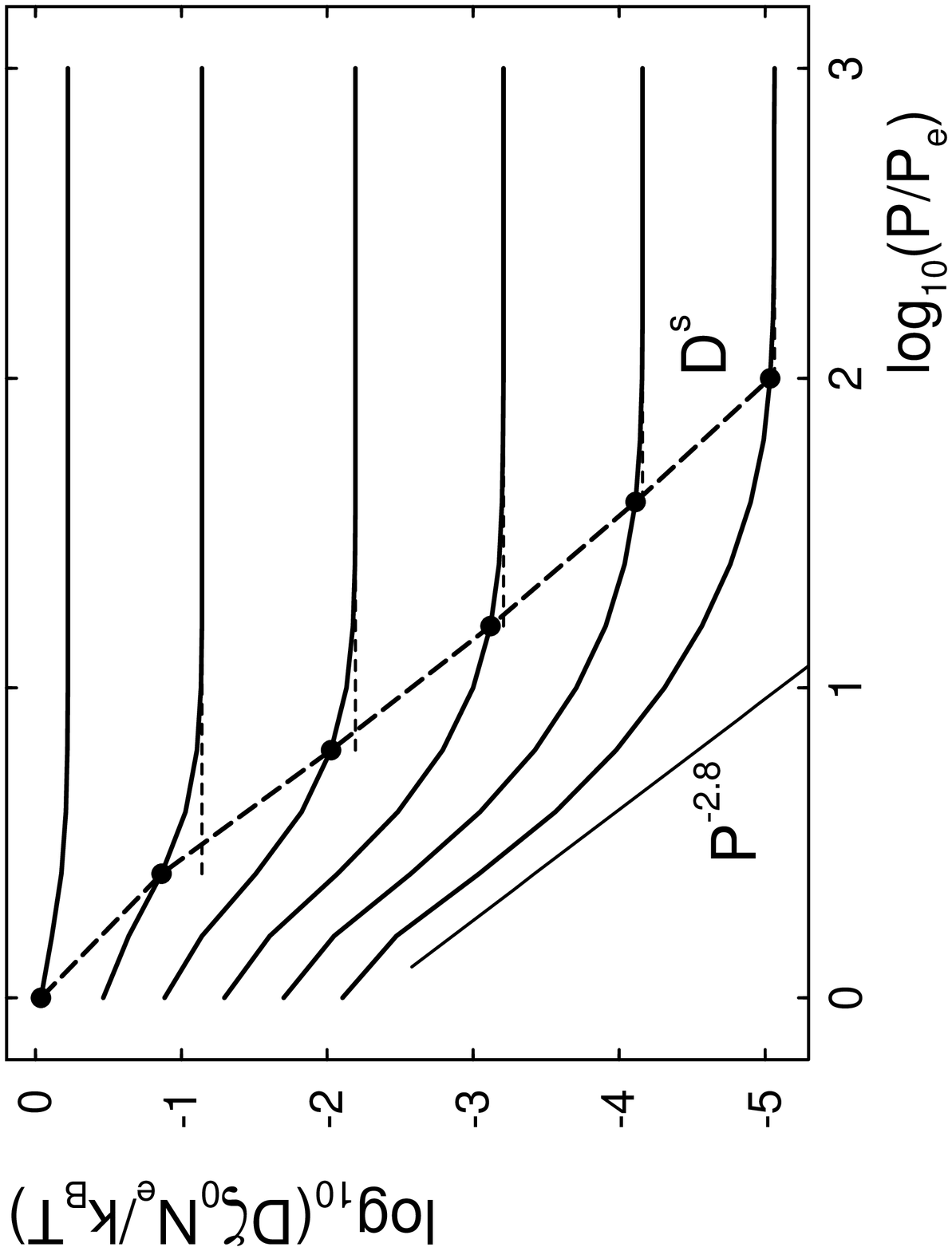}}}
\caption{ }\label{f5}
\end{figure}

\begin{figure}[H]
\centerline{\rotate[r]{\epsfysize=18.cm 
\epsffile{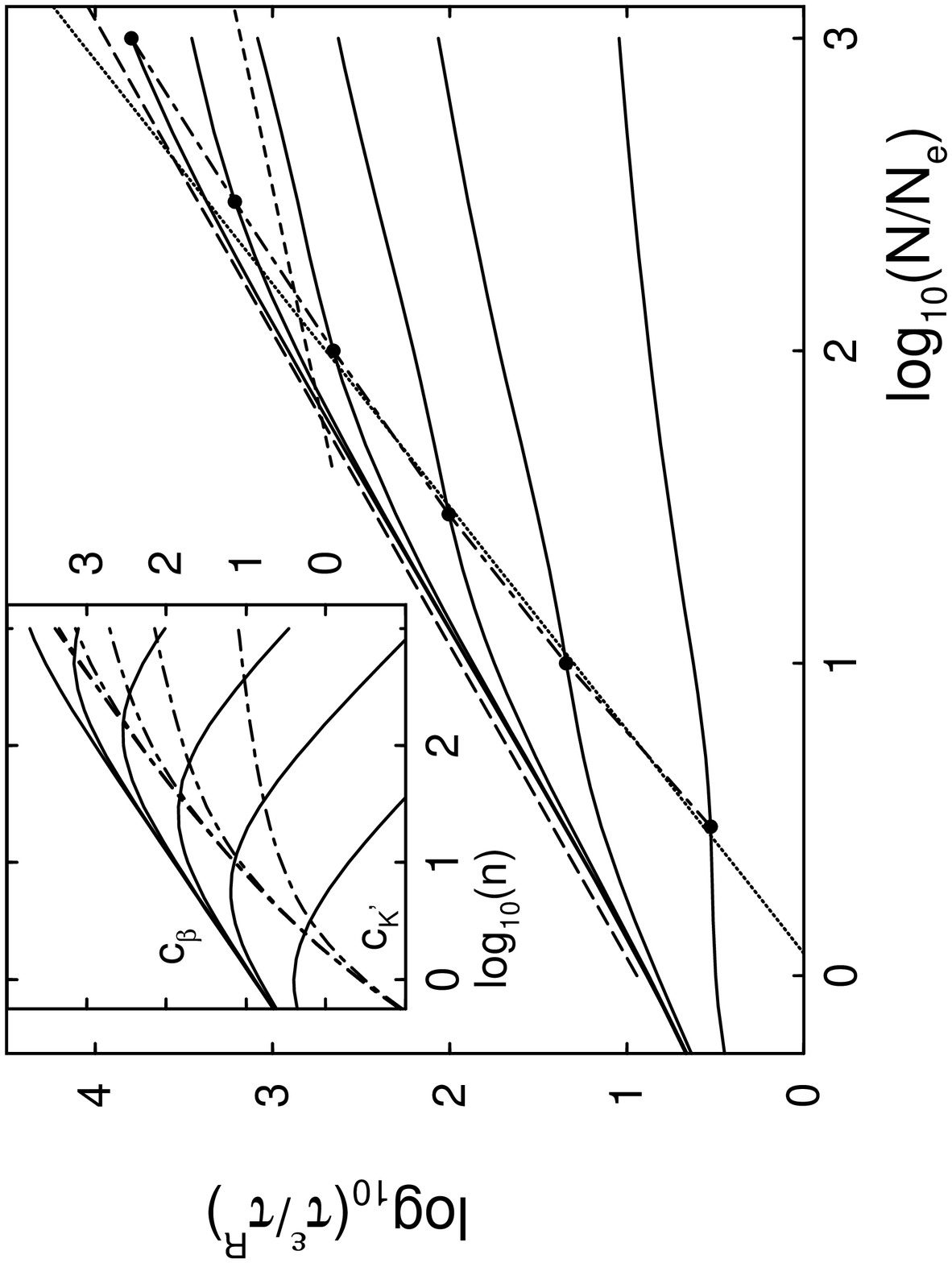}}}
\caption{ }\label{f6}
\end{figure}

\newpage
\bigskip
\bigskip
\leftline{\large{\bf Figure Captions}}
\begin{itemize}
\item{\bf Figure 1:}$\;$ 
Schematic representation of the (static) entanglement constraints in the
uniform--drag friction function, $\Sigma(t)$ eqn. (\protect\ref{14}), of  PMC
theory. 
The tracer, characterized by the intramolecular correlations
$\omega$  and length scales $\sigma$ and $R_g$, interacts via the 
short--ranged  (excluded--volume diameter $d$) pseudo--potential, $c$ (the 
direct correlation function) with the matrix polymers. The entanglement
constraints of the matrix separate into local (density
screening length $\xi_\rho$)  compressibility and mesoscopic 
 (entanglement length $b$) elastic correlations. 

\item{\bf Figure 2:}$\;$
Schematic figure showing the restriction of the
segment dynamics for times shorter than the disentanglement time. A discrete
bead--spring polymer model is used, where eqn. (\protect\ref{30}), with the
unknown $\bf c_\alpha$ set to $\bf c_\alpha=0$ for simplicity,
allows the determination
of all bead displacements from the motions, $\bf u(t)$  and
$\bf v(t)$, of two arbitrary segments.

\item{\bf Figure 3:}$\;$
Schematic representation of the (static) entanglement constraints in the
friction functions of  the PMC theory.
The tracer interacts via the short ranged pseudo--potential $c$ (the 
direct correlation function) with the matrix polymers. The entanglement
constraints of the matrix separate into local (density
screening length $\xi_\rho$)  compressibility (structure factor $S$)
and mesoscopic 
(entanglement length $b$) elastic correlations (amplitude $f$).
 Note that only a small number
of the  matrix polymers are shown as the blobs of size $\xi_\rho$ which
fill space. 
The tracer exhibits different
intramolecular  correlations, $I$, for the uniform drag, $I^\Sigma\sim\omega$,
and for the conformational friction, $I^M\sim \omega^2 / k^2$. Note that the
conformational friction coarse grains the matrix correlations over much larger
spatial regions.  

\item{\bf Figure 4:}$\;$ Intramolecular factors determining the entanglement
constraint amplitudes in the conformational, $I^M$ (long dashes) 
eqn. (\protect\ref{61}),
and in the center--of--mass, $I^\Sigma$ (solid line, scaled by $1/10$)
eqn. (\protect\ref{60}),
friction functions plotted versus reduced wavevector, $q=kR_g$.
The corresponding
asymptotic friction amplitudes, $\Gamma^\Sigma= 2 I^\Sigma\hat\omega/q^2$
(chain curve) and  $\Gamma^M= 2 I^M\hat\omega/q^2$ (short dashes),
 follow from the diffusive relaxation
rates,  $2 \hat \omega_q/ q^2$, of the RR model in eqn. (\protect\ref{32}).

\item{\bf Figure 5:}$\;$ Finite size reduction factors, $\beta(n,\alpha,\delta)
= \mtaud \alpha N_e / \tau_0 N^3$, of eqn. (\protect\ref{63}), for different
entanglement strengths, $\langle |F^2|\rangle/\varrho_m\sim 1/\alpha$, and
length scale ratios, $\delta = \xi_\rho/b$. The thin lines correspond to
the parameters $\alpha$ as denoted and to increasing parameter $\delta$ from
left to right.  The solid line shows the 
solution, $\beta(n/\alpha^2)$, neglecting constraint porosity, and the dotted
line the asymptote eqn. (\protect\ref{68}). 

\item{\bf Figure 6:}$\;$ Ratios of shear viscosity to Rouse
viscosity corresponding to the disentanglement times \taud following from the
results of figure \protect\ref{f1}. Bold lines denote the results neglecting
constraint porosity for different $\alpha$ and the thin lines for fixed
parameter  $\alpha$ shift with increasing  parameter $\delta$ from left to
right. Lines of the same style belong to one value of parameter $\alpha$. 
A power law, $\eta/\eta^R \sim N^{2.4}$, is shown for comparison. 
The $N\to\infty$ asymptote $\lambda_\eta (N/N_e)^2$ is drawn as a
dotted line for $\alpha=2$ or $\lambda_\eta=6$. 

\item{\bf Figure 7:}$\;$ Tracer diffusion constants in the limit of immobile
matrix polymers, $P\gg N$.
 Two sets of curves are shown with two different values of
the asymptotic prefactor, $\lambda_D$, and corresponding to upper or lower
horizontal scale. The curves show increasing steepness in the intermediate
region with increasing $\delta=\xi_\rho/b$;
 $\delta=$ 0.01, 0.05, 0.1, 0.2, 0.3 and 0.5
respectively.  Three  power laws with arbitrary prefactors are shown for
comparison.  

\item{\bf Figure 8:}$\;$ Tracer diffusion coefficients in a polymer melt 
as a function of matrix
molecular weight, $P/P_e$,  for different tracer degrees of polymerization,
$N/N_e$; $log_{10}(N/N_e)=$ 0, 0.4, 0.8, 1.2, 1.6 and 2 from top to bottom.   
The parameters employed,
 $\alpha=3$, $\delta=0.05$ and $\lambda_D=3$, are argued to
describe  a polymer melt. 
The self diffusion constants, $D^s$, 
are denoted by circles and long dashes, and can be
compared to the asymptotic, $D^{\rm tr}$ for  $P\to\infty$, 
values shown with horizontal short dashes. A
power $D\sim P^{-2}$ is shown for comparison.

\item{\bf Figure 9:}$\;$ 
 Tracer diffusion coefficients in a polymer solution as a function of matrix
molecular weight, $P/P_e$,  for different tracer degrees of polymerization,
$N/N_e$; $log_{10}(N/N_e)=$ 0, 0.4, 0.8, 1.2, 1.6 and 2 from top to bottom.   
The parameters employed,
 $\alpha=3$, $\delta=0.3$ and $\lambda_D=18$, are argued to
describe a polymer solution or a melt exhibiting
 a special tracer, matrix polymer interaction.
The self diffusion constants, $D^s$ (circles and long dashes),
 the asymptotic values, $D^{\rm tr}$ for $P\to\infty$ (short dashes), and a 
power $D\sim P^{-2.8}$, are shown. 

\item{\bf Figure 10:}$\;$ Tracer dielectric relaxation time,
$\tau^\varepsilon$,  normalized by the Rouse time as a function of reduced
tracer molecular weight, $n=N/N_e$, for different matrix molecular weights,
$p=P/P_e$; $p=$ 3, 10, 30, 100, 300, and 1000 from bottom to top. Melt like
parameters, $\alpha=3$ and $\delta=0.05$, are chosen.
Chemically identical tracer and matrix polymers are considered. 
The  asymptotes, $\tau^\varepsilon\sim N^3$ for $n\ll p$ (long dashes)
and $\tau^\varepsilon\sim N^{19/8}$ for $n\gg p$ (short dashes), are included.
The matrix end--to--end--vector relaxation times, shown with circles and a
dashed--dotted line, correspond to  $n=p$. A dotted line shows a power law,
$\tau^\varepsilon \sim N^{3.4}$, fitted through these $n=p$ system points. 
The inset separately shows the contributions in \gl{dielec} from the
conformational friction  corrections, $c_\beta=24 \beta n/\alpha$  (solid
lines), and from the homogeneous  friction, $c_{K'}=8 K' n/\alpha$  (chain
curves).    

\item{\bf Figure 11:}$\;$ Tracer diffusion constants normalized by the
Rouse result versus tracer molecular weight for tracer motion through gels with
different pore sizes; the correlation lengths $\xi_g$ agree with the radii of
gyration of tracers of the degrees of polymerization, $N/N_e=$ 10, 50, 100,
500, 1000. Tracer mass scaling exponent for a self avoiding walk polymer,
$\nu=3/5$, and gel fractal dimension, $D_F=5/3$, are used while fixing the
interaction strength to $\lambda_g=100$ in part (a) and $\lambda_g=20$ in
(b). For a soft, polymeric gel, eqn.  (\protect\ref{gelso}),
 the ratio of screening and entanglement lengths is fixed to
$\xi/b=$ 0.3, and thick solid lines are drawn. Chain curves correspond to   
the hard, structurally rigid  gel case, eqn. (\protect\ref{gelha}). The thin
line is the asymptote $DN\to N^{-(4-2\nu d)}$ which meets the curves for
$N/N_e\approx 10^8$.  The maximal intermediate slope, $DN\sim
N^{-(4-2\nu(d-D_F))}$, is shown as a dotted line. In part (a), power laws,
$DN\sim N^{-1.7}$ (short dashes) and $DN\sim N^{-1.5}$ (long dashes), are
compared to the soft and  hard gel calculation at $\xi_g=R_g(N=500N_e)$
respectively.
In part (b), power laws,
$DN\sim N^{-1.3}$ (short dashes) and $DN\sim N^{-1.2}$ (long dashes), are
compared to the soft and  hard gel calculation at $\xi_g=R_g(N=500N_e)$
respectively.

\end{itemize}

\newpage
\pagestyle{empty}

\newpage

\newpage

\newpage

\newpage

\newpage

\newpage

\newpage

\newpage

\newpage

\begin{figure}[H]
\centerline{\rotate[r]{\epsfysize=13.5cm 
\epsffile{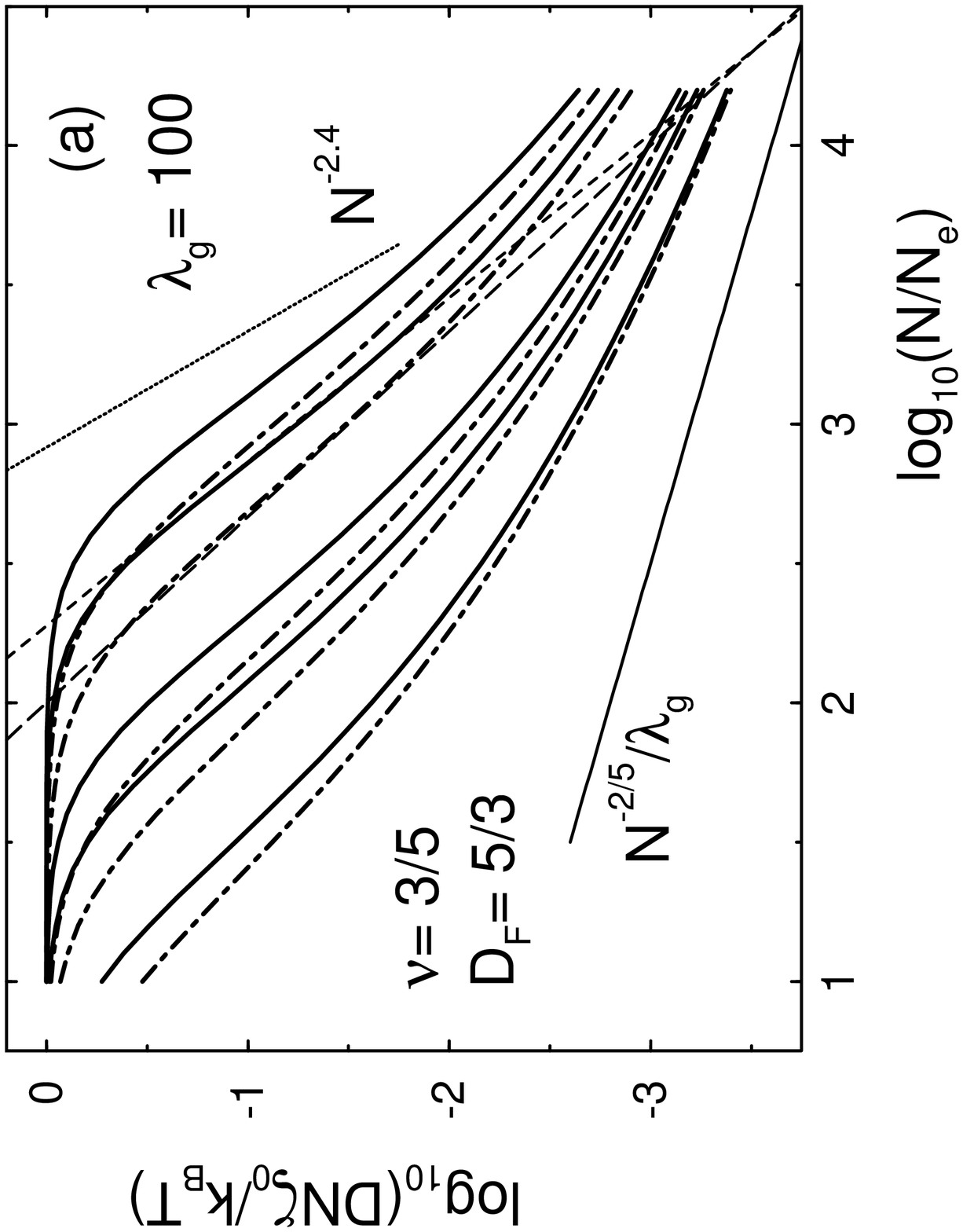}}}
\end{figure}
\begin{figure}[H]
\centerline{\rotate[r]{\epsfysize=13.5cm 
\epsffile{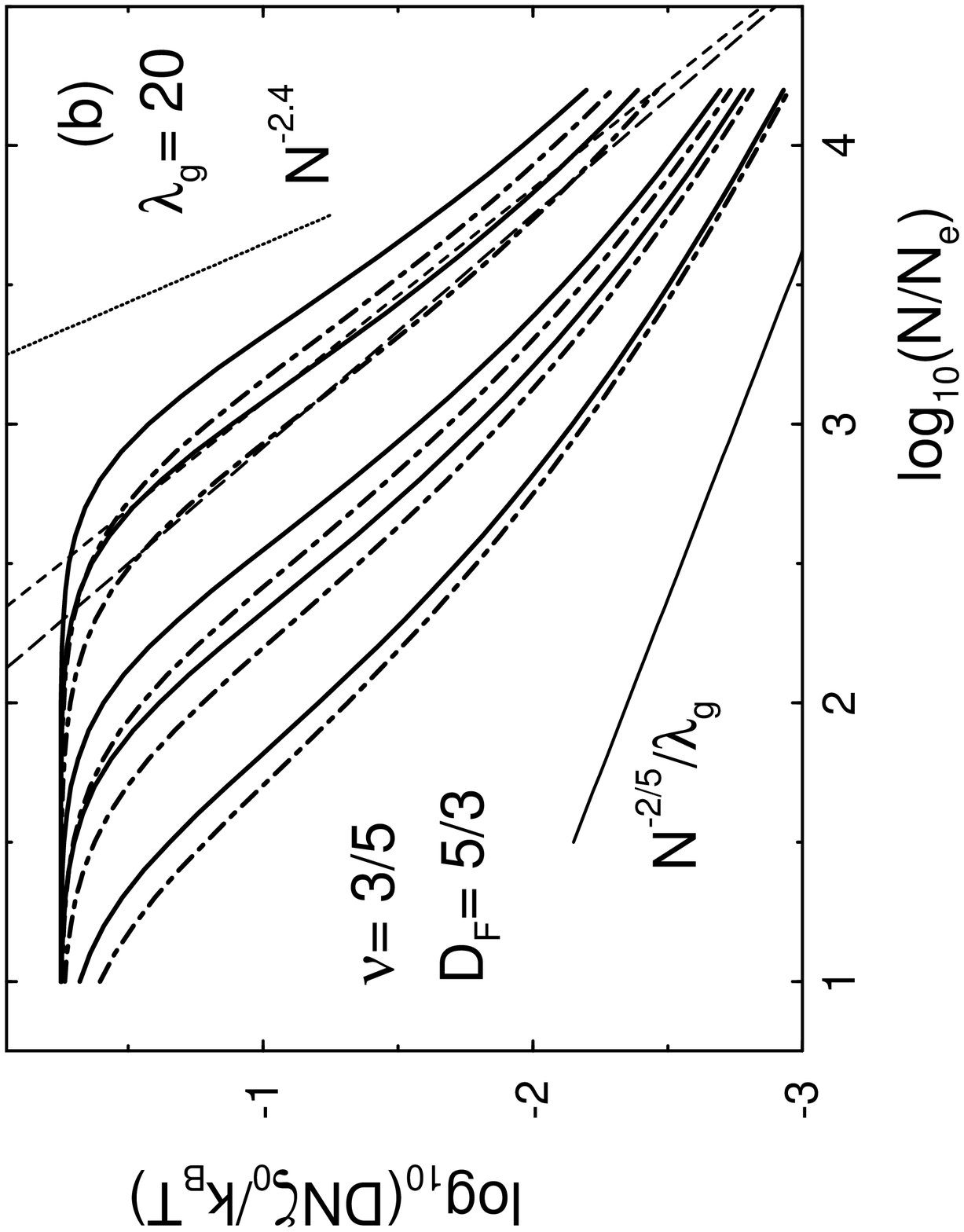}}}
\caption{ }\label{f7}
\end{figure}

\end{document}